\documentclass{aa}  
\usepackage{graphicx}
\usepackage{txfonts}
\usepackage{longtable}
\usepackage{natbib}
\bibpunct{(}{)}{;}{a}{}{,}
\def\sOri{{$\sigma$Orionis}}
\begin{document}
%
   \title{A census of very-low-mass stars and brown dwarfs in the $\sigma$ Orionis cluster \thanks{This work is based in part on data obtained as part of the UKIRT Infrared Deep Sky Survey. The United Kingdom Infrared Telescope is operated by the Joint Astronomy Centre on behalf of the Science and Technology Facilities Council of the U.K.}}

   \subtitle{}

   \author{N. Lodieu \inst{1,2}
          \and
          M. R. Zapatero Osorio \inst{1,2}
          \and
          R. Rebolo \inst{1,2}
          \and
          E. L. Mart\'{\i}n \inst{3,1,2}
          \and
          N. C. Hambly \inst{4}
           }

   \offprints{N. Lodieu}

   \institute{Instituto de Astrof\'isica de Canarias, C/ V\'ia L\'actea s/n,
                 E-38200 La Laguna, Tenerife, Spain\\
              \email{nlodieu@iac.es,mosorio@iac.es,rrl@iac.es,ege@iac.es}
              \and 
              Departamento de Astrof\'isica, Universidad de La Laguna, 
                 E-38205 La Laguna, Tenerife, Spain
              \and
             Centro de Astrobiolog\'ia (CSIC/INTA), 28850 Torrej\'on de Ardoz,
             Madrid, Spain
              \and
             Scottish Universities' Physics Alliance (SUPA),
             Institute for Astronomy, School of Physics \& Astronomy, University of Edinburgh,
             Royal Observatory, Blackford Hill, Edinburgh EH9 3HJ, UK \\
             \email{nch@roe.ac.uk}
             }

   \date{\today{}; \today{}}

 
  \abstract
   {The knowledge of the initial mass function (IMF) in open clusters 
constitutes one way of constraining the formation of low-mass stars and 
brown dwarfs, along with the frequency of multiple systems and the 
properties of disks.}
   {The aim of the project is to determine the shape of the mass function 
in the low-mass and substellar regimes in the $\sigma$ Orionis cluster 
($\sim$3 Myr, $\sim$352 pc, solar metallicity) as accurately as possible
and compare it with the results in other clusters.}
   {We have analysed the near-infrared photometric data from the Fourth
Data Release (DR4) of the UKIRT Infrared Deep Sky Suvey (UKIDSS) 
Galactic Clusters Survey (GCS) to derive the cluster luminosity and mass
functions, evaluate the extent of the cluster, and study the distribution 
and variability of low-mass stars and brown dwarfs down to the 
deuterium-burning limit.}
   {We have recovered most of the previously published members and found 
a total of 287 candidate members within the central 30 arcmin in the 
0.5--0.009 M$_{\odot}$ mass range, including new objects not previously
reported in the literature. This new catalogue represents a
homogeneous dataset of brown dwarf member candidates over the central
30 arcmin of the cluster. The expected photometric contamination by 
field objects with similar magnitudes and colours to $\sigma$ Orionis 
members is $\sim$ 15\,\%. We present evidence of variability at the 
99.5\% confidence level over $\sim$yearly timescales in 10 member 
candidates that exhibit signs of youth and the presence of disks. The level 
of variability is low ($\le$ 0.3 mag) 
and does not impact the derivation of the cluster luminosity and mass 
functions. Furthermore, we find a possible dearth of brown dwarfs within 
the central five arcmin of the cluster, which is not caused by a lower 
level of photometric sensitivity around the massive, O-type multiple star 
$\sigma$ Ori in the GCS survey. Using state-of-the-art theoretical models, 
we derived the luminosity and mass functions within the central 
30 arcmin from the cluster centre, with completeness down to 
$J$ = 19 mag, corresponding to masses ranging from 0.5 M$_{\odot}$ down 
to the deuterium-burning mass boundary ($\sim$0.013 M$_{\odot}$).}
   {The mass function of \sOri{} in this mass interval shows a power law 
index $\alpha$ = 0.5$\pm$0.2 (when expressed as 
dN/dM$\propto$M$^{-\alpha}$), which agrees with the one 
derived for the 3--5 Myr cluster Upper Sco (based on similar data obtained 
with the GCS) in the same mass range.
}


   \maketitle
%

%
%
\section{Introduction}

The number of objects per unit of mass, known as the initial mass
function (hereafter IMF), is of prime importance in understanding
stellar and substellar formation processes. After the pioneering studies 
by \citet{salpeter55}, \citet{miller79}, and \citet{scalo86},
the stellar IMF has been investigated during the past decades in 
various environments and over a wide mass range to look into a
possible dependence on time and place. To address some of the crucial 
questions like the universality of the IMF, young open clusters and 
star-forming regions have been extensively targeted because they represent 
a coeval population of stars of similar metallicity at a given distance.
Results from large-scale surveys of open clusters suggest that no 
clear and unambiguous variation in the IMF has been seen and that the
IMF is consistent with the field mass function \citep{kroupa02,chabrier03}.
However, \citet{luhman03b} find evidence of a clear variation in
the IMF between Taurus that peaks at 0.8 M$_{\odot}$ and other young
regions (e.g.\ IC\,348; peak at 0.1--0.2 M$_{\odot}$).

The UKIRT Deep Infrared Sky Survey \citep[UKIDSS;][]{lawrence07}, 
made of five sub-surveys, is designed to reach three to four magnitudes 
deeper than 2MASS \citep{cutri03} and cover several thousands of square 
degrees in several infrared filters. The UKIDSS project is described in 
\citet{lawrence07} and uses the Wide Field Camera \citep[WFCAM;][]{casali07} 
installed on the UK InfraRed Telescope (UKIRT) and the Mauna Kea
Observatory \citep{tokunaga02} photometric system described in 
\citet{hewett06}. The pipeline processing is described in 
Irwin et al.\ (in prep)\footnote{Extensive details on the data
reduction of the UKIDSS images is available at
http://casu.ast.cam.ac.uk/surveys-projects/wfcam/technical}
and the WFCAM Science Archive (WSA) in \citet{hambly08}.

The Galactic Clusters Survey (GCS), one of the UKIDSS components,
aims to make a census of young brown dwarfs in ten star-forming regions and open
clusters over large areas in five passbands ($ZYJHK$) across the
1.0--2.5 micron wavelength range with a second epoch in $K$.
The main goal is to measure the form of the IMF
\citep{salpeter55,miller79,scalo86} in the low-mass and substellar 
regimes to investigate important issues, including the formation
and spatial distribution of low-mass stars and brown dwarfs.
Early results on the IMF in the Pleiades and in Upper Scorpius are 
presented in \citet{lodieu07c} and \citet{lodieu07a}, respectively,
and are used for comparison with \sOri{} in this paper.

The $\sigma$\,Orionis cluster, located around the O9.5-type multiple star 
of the same 
name, belongs to the Orion OB 1b association and was first mentioned
by \citet{garrison67} and later by \citet{lynga81}. The X-ray detection 
of a high concentration of sources around the central star by ROSAT
\citep{walter94} triggered deep optical surveys dedicated to the
search for low-mass stars and brown dwarfs. The cluster is 1--8\,Myr old
\citep{bejar99} with a most probable age of 3$\pm$2 Myr 
\citep{oliveira02,zapatero02a,sherry04} and suffers from little reddening 
\citep{lee68,bejar99}. The cluster is located at 352$^{+166}_{-85}$ pc 
according to {\it {Hipparcos}} \citep{perryman97} but could actually be 
further away, at a distance of 440 pc \citep{brown94,sherry04,jeffries06}.
Deep optical surveys of a large area of the cluster complemented 
by near-infrared photometry revealed numerous low-mass stars, brown 
dwarfs \citep{bejar99,kenyon05}, and planetary-mass members
\citep{zapatero00}. Many objects have been spectroscopically confirmed 
over a large mass range in the optical 
\citep{bejar99,zapatero00,bejar01,barrado01c,kenyon05,sacco08} 
and in the near-infrared \citep{zapatero00,martin01a}.
Many sources have now mid-infrared counterparts with the advent of
the {\it{Spitzer}} space telescope 
\citep{hernandez07,caballero07d,zapatero07b,scholz08a,luhman08c}, 
allowing an estimate of the disk fraction through the mid-infrared
excesses. The cluster mass function, derived across the low-mass 
($\leq$0.2 M$_{\odot}$), substellar, and planetary-mass regimes, 
indicates a rising slope with an index $\alpha$ = 0.8$\pm$0.4 
\citep{bejar01}, recently revised by \citet{caballero07d} to 
$\alpha$ = 0.6$\pm$0.2 (when expressed as the mass spectrum, 
dN/dM$\propto$M$^{-\alpha}$).
Moreover, \citet{jeffries06} reported the presence of two distinct 
populations in \sOri{} but indistinguishable in colour-magnitude 
diagrams (see also \citet{caballero07b}). 
The first one is associated with \sOri{} itself and has an age around 
3 Myr and a distance of 440 pc whereas the second group appears older 
($\sim$10 Myr) and closer (d $\sim$ 352 pc).

The GCS provides a full near-infrared coverage of \sOri{} in five filters
over the central 30 arcmin (Fig.\ \ref{fig_sOri_GCS:coverage}). In this 
paper we present a census of very low-mass star (referred to as 
M $\leq$ 0.5 M$_{\odot}$ throughout this paper) and 
brown dwarf member candidates down to the deuterium-burning limit 
($J \sim$ 17.9 mag) in \sOri{}. In Sect.\ \ref{sOri_GCS:selection} we 
describe the selection of photometric and proper motion member candidates. 
In Sect.\ \ref{sOri_GCS:compare_cat}, we compare our catalogue with 
previous studies of low-mass stars and brown dwarfs identified in \sOri{}.
In Sect.\ \ref{sOri_GCS:variability}, we explore the variability of
cluster member candidates. In Sect.\ \ref{sOri_GCS:extent} we briefly 
evaluate the extent of \sOri{} on the sky.
In Sect.\ \ref{sOri_GCS:distribution}, we investigate the distribution
of low-mass stars and brown dwarfs across the central 30 arcmin of the
cluster. In Sect.\ \ref{sOri_GCS:contamination} we discuss the level of 
contamination affecting our photometric selection across the full magnitude 
range. Finally, we derive the cluster luminosity and mass functions 
for the central 30 arcmin down to the deuterium-burning limit and compare 
our results to earlier studies in \sOri{} and other clusters
(Sect.\ \ref{sOri_GCS:MF_LF}).

%
%
\section{The selection of cluster member candidates}
\label{sOri_GCS:selection}
\subsection{The query}
\label{sOri_GCS:selection_SQL}

\begin{figure}
   \centering
   \includegraphics[width=\linewidth]{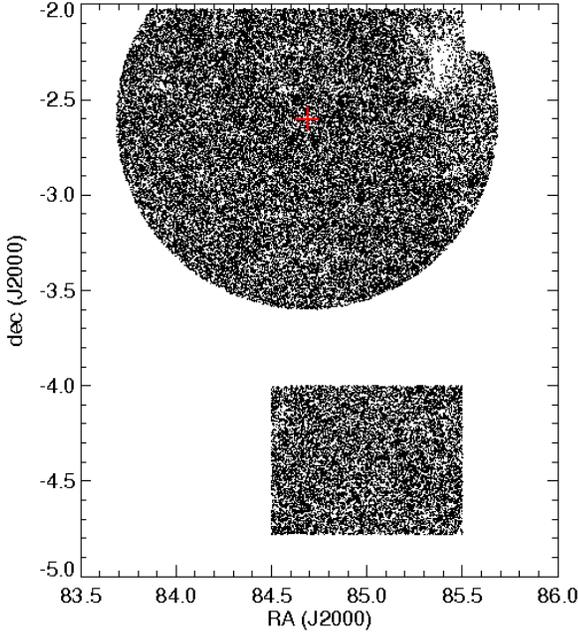}
   \caption{Coverage of the UKIDSS GCS in \sOri{}. All sources in the
GCS database within a radius of 60 arcmin from the central star, $\sigma$
Ori, are plotted along with the control field used to estimate the
photometric contamination.
}
   \label{fig_sOri_GCS:coverage}
\end{figure}

We extracted from the UKIDSS GCS DR4 all point sources located
within a one degree radius from the star $\sigma$ Ori which marks the
centre of the cluster (Fig.\ \ref{fig_sOri_GCS:coverage}). The
survey coverage is spatially homogeneous and complete up to $\sim$35
arcmin because the part north of Declination equal to $-$2 deg is
missing in GCS DR4\@. The central coordinates of the tiles observed
by the GCS in \sOri{} are listed in Table \ref{tab_sOri_GCS:log_obs}. 
Each WFCAM tile (54 by 54 arcmin aside) was observed in all five
passbands ($ZYJHK$). Average seeing measured on the \sOri{} images and in 
each individual filter ranged from 0.75 to 1.0 arcsec.
Central coordinates (in J2000) of each WFCAM tile
along with the corresponding observing date are provided in the table.

\begin{figure}
   \centering
   \includegraphics[width=\linewidth]{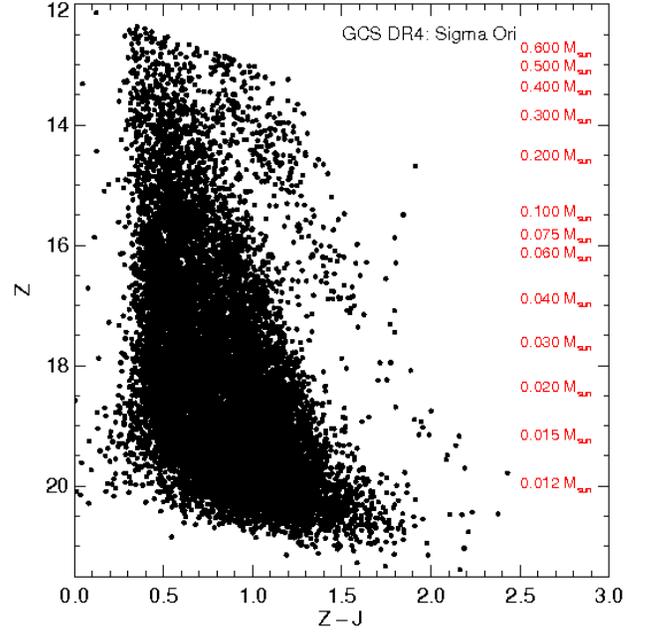}
   \caption{($Z-J$,$Z$) colour-magnitude diagram
for the full area within 30 arcmin of the centre of the \sOri{} cluster,
represented by the sigma Ori star itself. The mass scale is indicated
on the right-hand side, using the NextGen \citep{baraffe98} and DUSTY
\citep{chabrier00c} models and assuming a distance of 352 pc and an age
of 3 Myr for \sOri{}.
}
   \label{fig_sOri_GCS:cmd_zjz_alone}
\end{figure}

We used a similar Structured Query Language (SQL) query to our work 
in Upper Sco \citep{lodieu06,lodieu07a} and the Pleiades \citep{lodieu07c}
to extract the catalogue of sources towards \sOri{}. Briefly, 
we selected only point sources
({\tt{Class}} parameter equal to $-$2 or $-$1) detected in $ZYJHK$ as
well as those detected only in $JHK$. The resulting catalogue contained
46884 sources. All sources within 30 arcmin (to guarantee a
full spatial coverage, and also see Sect.\ \ref{sOri_GCS:extent})
are plotted in the ($Z-J$,$Z$) colour-magnitude diagram presented in
Fig.\ \ref{fig_sOri_GCS:cmd_zjz_alone}. We estimate that the survey
towards \sOri{} is complete in the following magnitude ranges: 12.2
$\leq Z \leq$ 20.2 mag, 12.4 $\leq Y \leq$ 20.0 mag, 12.0 $\leq J
\leq$ 19.0 mag, 12.5 $\leq H \leq$ 18.4 mag, and 11 $\leq K \leq$ 18
mag. These magnitudes are estimated from the values where the
histograms of the numbers of sources as a function of magnitude
deviate from a power law fit to the counts.  Our analysis will
solely focus on candidates fainter than $J \sim$ 12 mag, corresponding
to low-mass star (M $\leq$ 0.5 M$_{\odot}$) and brown dwarf members
down to the deuterium-burning limit. The GCS provides five-band
photometry and our SQL query included the computation of the proper
motions from the cross-match between the 2MASS \citep{cutri03} and the
GCS for all sources brighter than $J$ = 15.5 mag.

%
%
%

%
%
\begin{table}
 \centering
  \caption{Log of the observations.}
 \label{tab_sOri_GCS:log_obs}
 \begin{tabular}{c c c c c c}
 \hline
Tile$^{a}$  &  R.A.  & Dec & Date       \\
      &   h:m   &  d:'  & YYYY-MM-DD \\
 \hline
1 &  05:33   & $-$03:21 & 2005-10-05  \\
2 &  05:33   & $-$02:28 & 2005-10-14  \\
3 &  05:36   & $-$03:21 & 2005-10-13  \\
4 &  05:36   & $-$02:28 & 2005-10-06  \\
5 &  05:39   & $-$03:21 & 2005-10-13  \\
6 &  05:40   & $-$02:28 & 2005-10-14  \\
7 &  05:41   & $-$03:21 & 2005-10-06  \\
8 &  05:43   & $-$02:28 & 2005-10-10  \\
9 &  05:45   & $-$03:21 & 2005-10-10  \\
 \hline
 \end{tabular}
\end{table}
%

%
%
\subsection{Selection in colour-magnitude diagrams}
\label{sOri_GCS:selection_CMDs}

The selection of cluster member candidates in \sOri{} was carried out
in a similar manner as the procedure outlined for the similarly young
Upper Sco \citep{lodieu06,lodieu07a} and the older Pleiades clusters
\citep{lodieu07c}. Nonetheless, we used the Mayrit catalogue
\citep{caballero08c} along with other spectroscopic known members (see
Section \ref{sOri_GCS:compare_cat}) to design photometric cuts in
three different colour-magnitude diagrams (Fig.\
\ref{fig_sOri_GCS:cmds_vpd}).

We restricted our selection to sources fainter than $J$ = 12 mag
and to the central 30 arcmin in radius around the massive
multiple star $\sigma$ Ori to avoid contamination by the Orion
complex (see Sect.\ \ref{sOri_GCS:extent}). We also defined a
combination of lines in three colour-magnitude diagrams to
separate the location of potential cluster members from foreground and
background field objects. These ``separators'', shown in Fig.\
\ref{fig_sOri_GCS:cmds_vpd}, are as follows:
\begin{itemize}
\item[$\bullet$] ($Z-J$,$Z$) from (0.6,12.0) to (1.2,16.5)
\item[$\bullet$] ($Z-J$,$Z$) from (1.2,16.5) to (2.1,21.5)
\item[$\bullet$] ($Y-J$,$Y$) from (0.40,12.0) to (0.65,17.0)
\item[$\bullet$] ($Y-J$,$Y$) from (0.65,17.0) to (1.10,20.5)
\item[$\bullet$] $J-K \geq$ 0.75 mag for $J$ = 12--16 mag 
\item[$\bullet$] ($J-K$,$J$) = (0.75,16.0) to (1.50,20.0)
\end{itemize}
These photometric cuts were guided by known bright members published by 
\citet{caballero08c} and faint spectroscopic members
\citep{bejar99,barrado01c,martin01a}. Cluster candidates were selected as 
those objects redder than the ``separators'' (dot-dashed lines) of 
Fig.\ \ref{fig_sOri_GCS:cmds_vpd}. The first photometric selection based
on the $Z-J$ colour returned 327 sources while the $Y-J$ colour gave
303 objects. The total number of candidates was 295 after
considering all three colors $Z-J$, $Y-J$, and $J-K$; this included
88 potential brown dwarfs with $J$ magnitudes fainter than $\sim$14.4
mag, assuming an age of 3 Myr and a distance of 352 pc
\citep{baraffe98} for the cluster. If we assume a larger distance of
440 pc, the stellar/substellar boundary would be at $J \sim$ 14.9 mag,
decreasing the number of brown dwarf candidates to 64. The selection
of candidates is complete down to $J$ = 19.0 mag, translating into a
mass of $\sim$0.0075 and $\sim$0.009 M$_{\odot}$ for a distance of 352
pc and 440 pc, respectively.

\subsection{Proper motions}
\label{sOri_GCS:selection_PM}

\begin{figure*}
   \centering
   \includegraphics[width=0.49\linewidth]{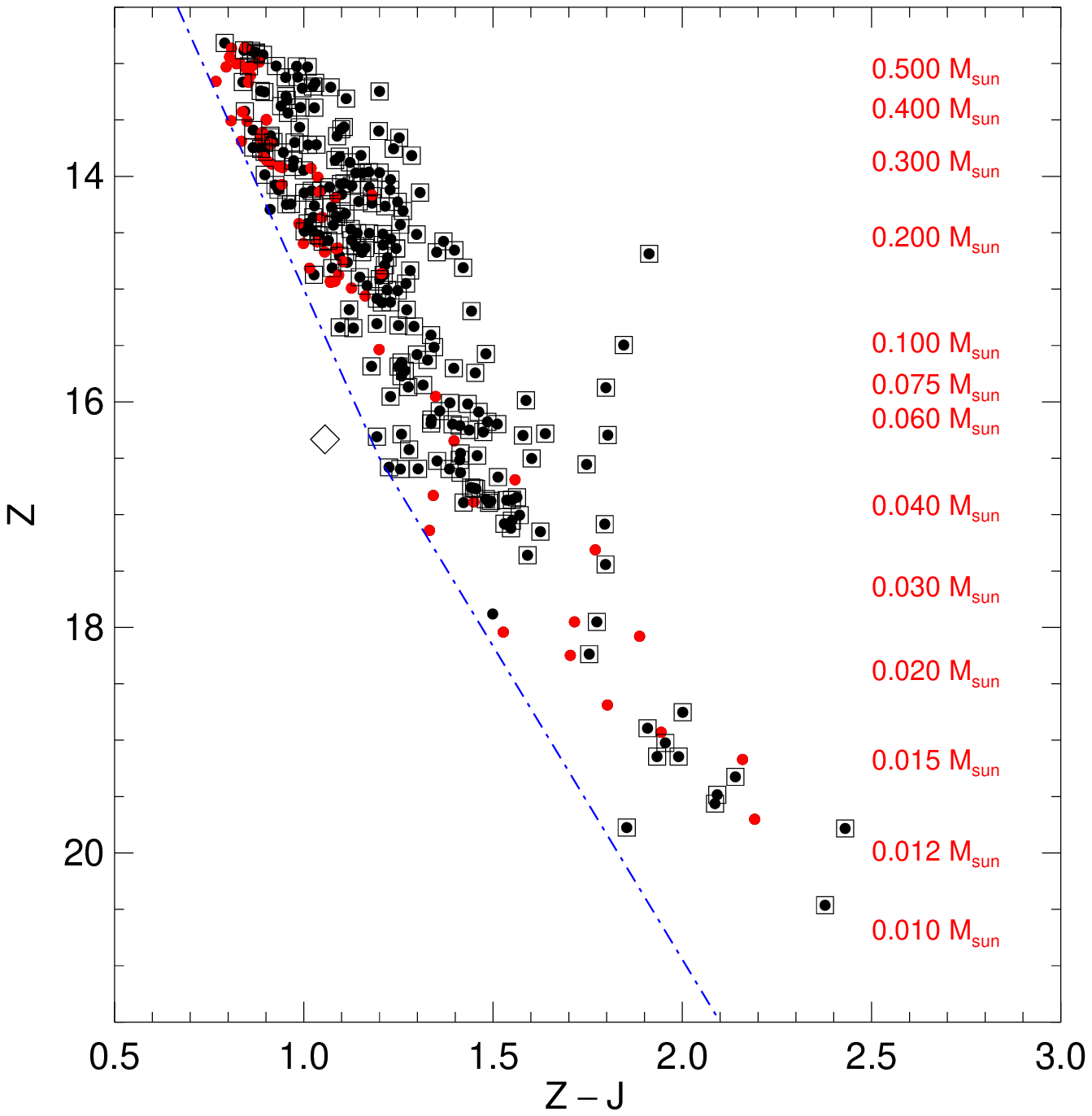}
   \includegraphics[width=0.49\linewidth]{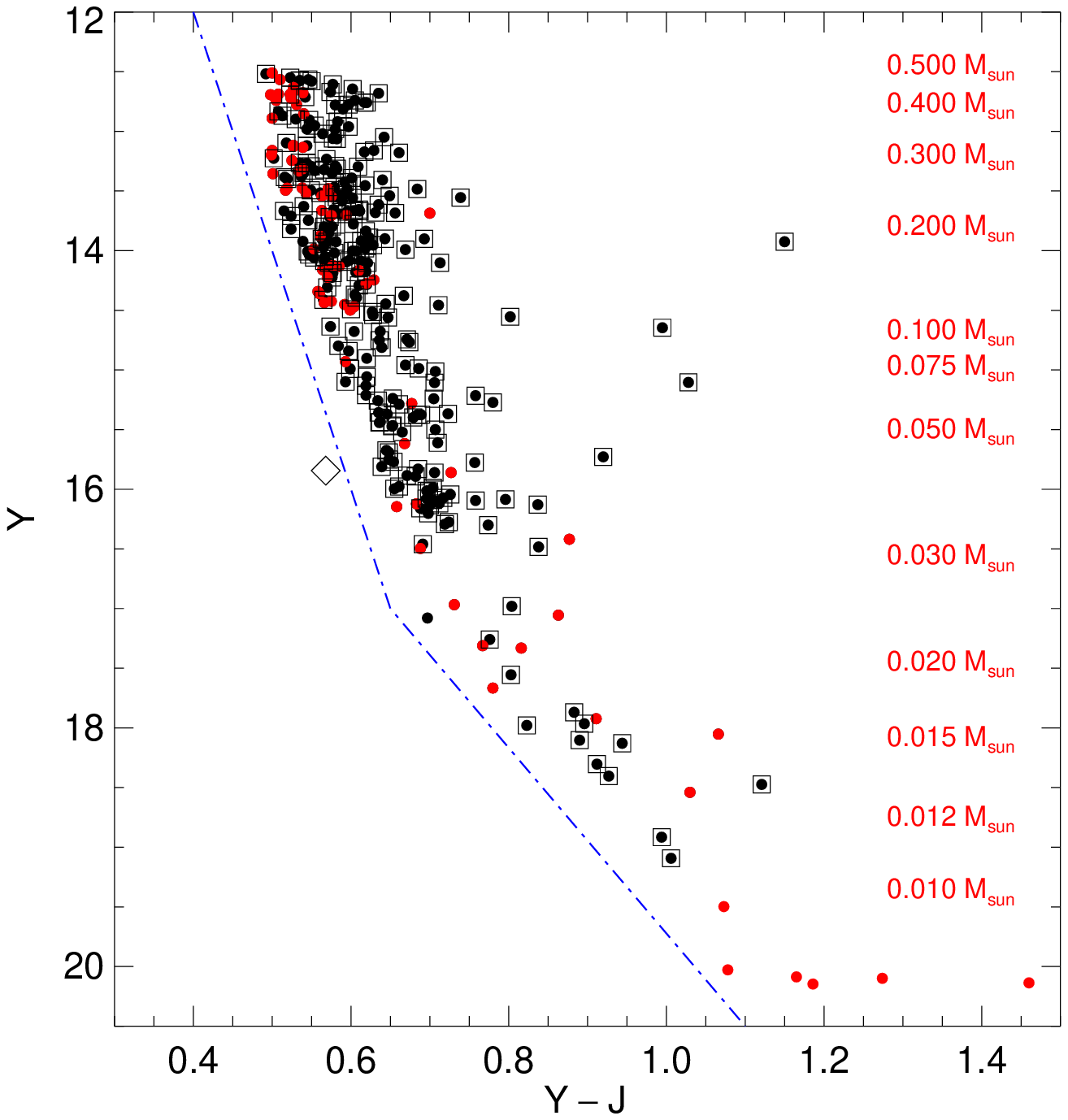}
   \includegraphics[width=0.49\linewidth]{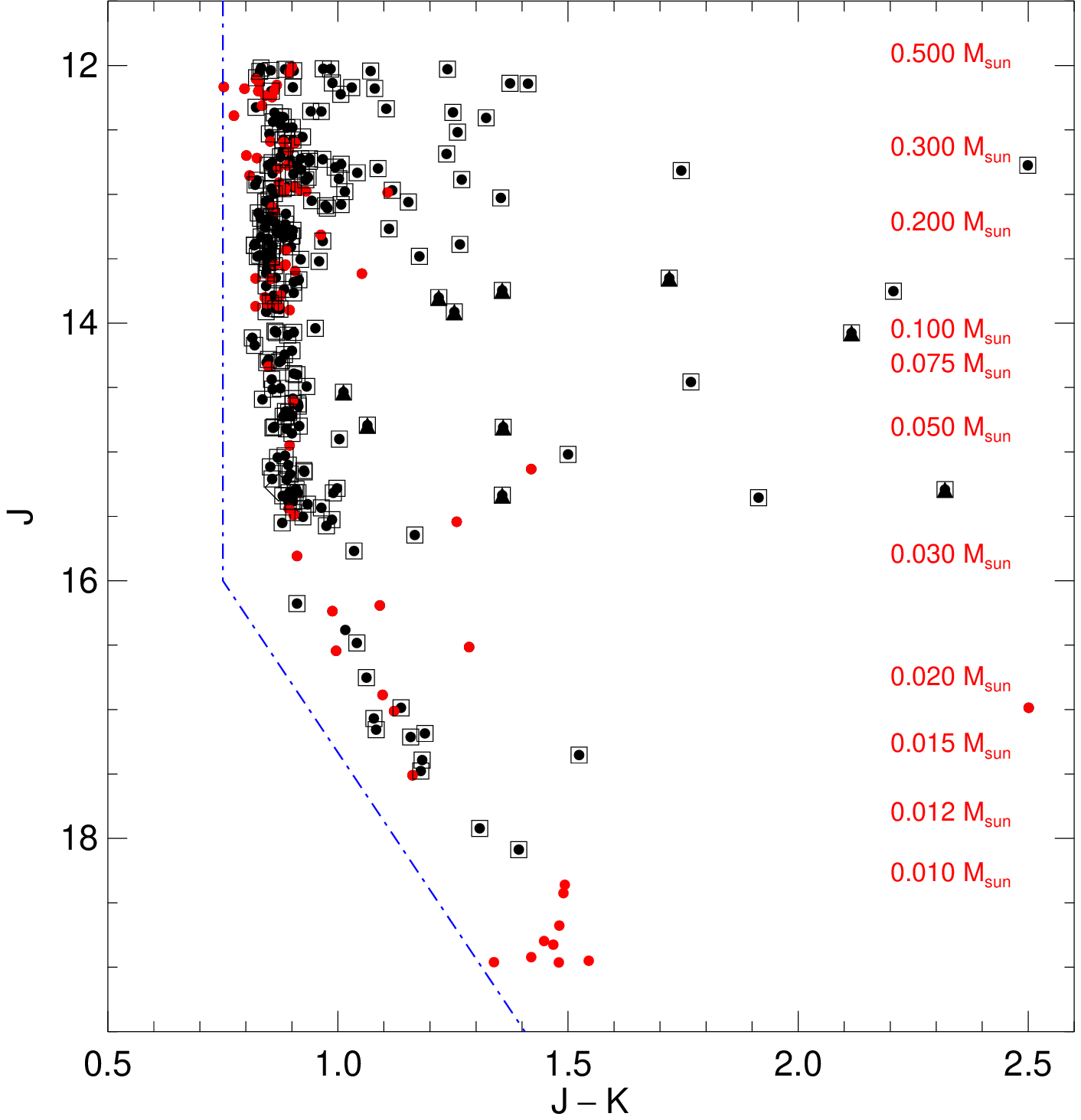}
   \includegraphics[width=0.49\linewidth]{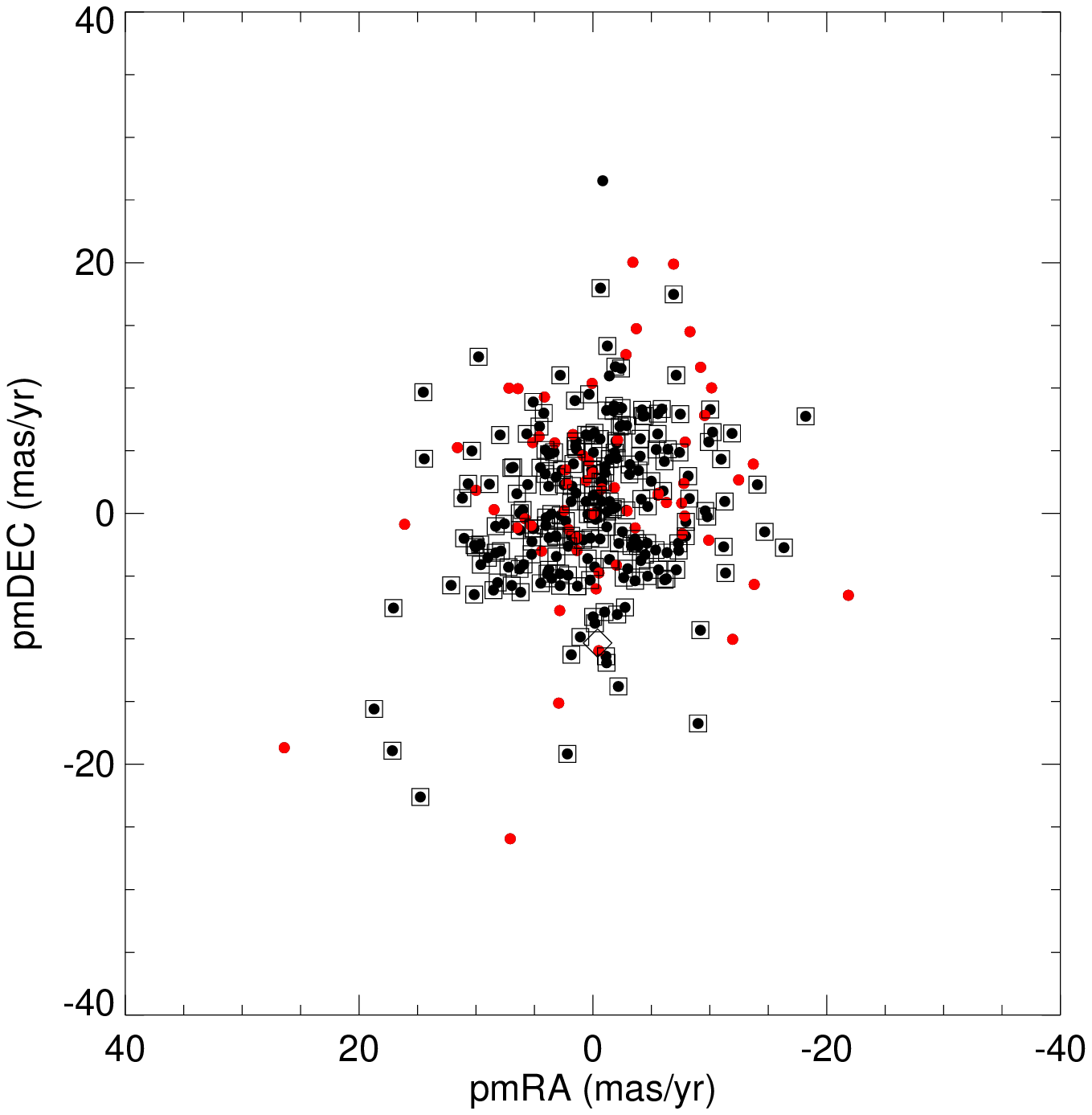}
   \caption{($Z-J$,$Z$), ($Y-J$,$Y$) and ($J-K$,$J$) colour-magnitude 
diagrams showing the sequences defined by the candidates identified 
in the GCS DR4 (dots) within the central 30 arcmin of \sOri{}. 
Known members of \sOri{} are shown as boxed points. Red symbols are
new photometric candidates identified in the GCS\@. 
The large diamond depicts the location of SE\,70 whose membership
is discussed in Sect.\ \ref{sOri_GCS:compare_cat_other} (difficult
to see in the ($J-K$,$J$) diagram at $J-K$ = 0.87 mag).
The filled triangles in the ($J-K$,$J$) diagram refer to variable
sources listed in Table \ref{tab_sOri_GCS:GCScand_variable}.
The mass scale is indicated on the right-hand side (for details, see the text).
Dot-dashed lines represent our selection criteria 
(Sect.\ \ref{sOri_GCS:selection_CMDs}). The bottom right plot show the 
vector point diagram for all candidate members with proper motion
measurement (three objects lie outside the vector point diagram).
}
   \label{fig_sOri_GCS:cmds_vpd}
\end{figure*}

Among the 293 sources extracted from the photometric selection, 263 are
brighter than $J$ = 15.5 mag and have proper motions accurate to 10 mas/yr
from the 2MASS/GCS cross-match implemented in our SQL query
(bottom right panel of Fig.\ \ref{fig_sOri_GCS:cmds_vpd}). The 2MASS
data in  \sOri{} were taken between October 1998 and November 2000,
yielding a baseline of 5--7 years between the 2MASS and GCS data. The mean 
proper motion of \sOri{} ($\mu_{\alpha}\cos{\delta}$,~$\mu_{\delta}$) 
= ($+$3.42$\pm$1.05,$-$0.20$\pm$0.60) mas/yr \citep{kharchenko05a} is 
not large enough to use proper motion
to separate potential cluster members from field star non--members as in the 
Pleiades \citep[e.g.][]{lodieu07c}. However, we can reject high proper motion 
sources with a proper motion 5--10$\sigma$ above the mean of the cluster 
\citep[$>$30 mas/yr;][]{caballero07b} taking into account the uncertainty
of 10 mas/yr in the accuracy of the proper motions. This lower limit allowed
us to identify eight objects: five are proper motion non-members and
three others are resolved close visual binaries in the UKIDSS images 
but were not on the
2MASS images, leading to an erroneous proper motion estimate
(Table \ref{tab_sOri_GCS:GCScand_PM_NM}). We do not include the
three potential binaries in our final catalogue because their
physical association should be confirmed first. The five sources
with large proper motions might be of interest for subsequent
astrometric follow-up to obtain their three dimensional motions and 
assess their membership to find out if they could have been ejected
from the cluster early on during the formation process or if they
could belong to a moving group associated with Orion.
To summarise, our $ZYJHK$ photometric and proper motion sample contains 
285 candidates.

\subsection{Faint candidates}
\label{sOri_GCS:selection_YJ}

To extend the cluster sequence to fainter magnitudes 
($J$ = 18--19 mag), we carried out a search for sources
detected in $YJHK$ but not in $Z$ using the same $Y-J$ and $J-K$
colour criteria as in Sect.\ \ref{sOri_GCS:selection_CMDs}.  Again, we
limited the search to the central 30 arcmin.  This query returned
10 sources in the magnitude interval $J$ = 18--19 mag but one
object lies at the edge of the detector and another one has a marginal
enhancement above the background in the $Z$-band but is
uncatalogued in the GCS source $Z$ detection table. The
cross-match between the sample of 10 candidates that lack $Z$-band
detection and the previously known members leads to the identification
of SOri\,56 and SOri\,60 \citep{zapatero00}. Only two sources are
brighter than $J$ =18.5 mag, SOri\,56 \citep{zapatero00} and SOri
J053829.5$-$025959, the brightest of all.  The errors on the $Y$
photometry are on the order of 0.2 mag while the errors on $J$ and $K$
are around 0.1 mag or below. The other five objects are new. We
provide $YJHK$ photometry with the associated errors for these 10
sources in Table \ref{tab_sOri_GCS:new_FAINT}.

We repeated the same search but for objects detected only in $JHK$
i.e., no detection in $Z$ and $Y$. We found five new candidates 
brighter than $J$ = 19 mag (Table \ref{tab_sOri_GCS:new_FAINT}). The two
brightest candidates lie very close to a bright star and are actually seen
on the $Z$ and $Y$ images but not catalogued in the GCS source detection
table likely due to the proximity of the bright star. The 
cross-match with deep Subaru $I$,$z$ data (unpublished) within 
3 arcsec shows that two sources are detected at $I$ and $z$ and exhibit 
blue optical ($I-z \sim$ 0.3--0.4 mag) and optical-to-infrared 
($z-J \sim$ 1.4--1.5 mag) colours inconsistent with cluster 
membership\footnote{The offsets between $z$-band filters is 
$z_{\rm Subaru}$ = $Z_{\rm GCS} \pm$0.5 mag}.
Therefore, only one source remains as a potential member from the
$JHK$ selection, SOri J053857.52$-$022905.5\@.

%
%
%

%
%
\section{Comparison with previous catalogues}
\label{sOri_GCS:compare_cat}

In this section, we discuss the cross-match of our list of
candidates identified in the GCS (filled circles in 
Fig.\ \ref{fig_sOri_GCS:cmds_vpd}) with previous cluster members 
published in the literature (squares in Fig.\ \ref{fig_sOri_GCS:cmds_vpd}), 
including the work at optical and near-infrared wavelengths by 
\citet{bejar99}, \citet{zapatero00}, \citet{bejar01}\footnote{$I$,$Z$
photometry of objects in the area survey surveyed by the Isaac Newton
Telescope Wide-Field camera can be found in the PhD thesis of Victor S\'anchez
B\'ejar \citep[University of La Laguna, Tenerife, Spain;][]{bejar01a,bejar01_PhD}}, 
\citet{kenyon05}, and \citet{burningham05a}
as well as the Mayrit catalogue \citep{caballero08c}. We have also
compared our list of candidates with the mid-infrared catalogues
of \citet{hernandez07} and \citet{luhman08c}.

\subsection{The Mayrit catalogue}
\label{sOri_GCS:compare_cat_Mayrit}

The Mayrit catalogue contains 338 sources in the $J_{\rm 2MASS}$ =
4.3--16.4 mag range, including 241 members showing at least one
youth feature described in \citet{caballero08c} and 97 member
candidates with no previous membership information. Among these 338
objects, 116 are brighter than $J_{\rm GCS}$ = 12 mag, the upper 
limit used for the selection of candidates in the GCS, 16 are not 
in the GCS database (9 of them have youth features) because they 
do not satisfy the criteria 
imposed on the {\tt{Class}} parameters in at least one passband 
(Sect.\ \ref{sOri_GCS:selection_SQL}), and the remaining 206
have a GCS counterpart within 2 arcsec. After cross-matching these
206 sources with our list of $\sigma$ Orionis candidate members
(Table \ref{tab_sOri_GCS:Known_Memb}), 198 of them nicely define 
the photometric sequence of the cluster in the colour-magnitude
diagrams depicted in Fig.\ \ref{fig_sOri_GCS:cmds_vpd}.
We note that four of the 198 sources 
fitting the cluster sequence are spectroscopic binaries
\citep[M258337, M459224, M873229, M1493050;][]{maxted08}. The other
sources listed in Table 4 of \citet{maxted08} are either photometric
and/or proper motion non-members.

Among the 206 sources common to the GCS and the Mayrit catalogues
\citep{caballero08c}, eight did not pass our selection photometric
criteria although the majority of those show signatures of a
young age. One source, M500279, is possibly a Class II object with
X-ray emission \citep{caballero07c,hernandez07} but has no $Y$
photometry for an unknown reason which leaves it outside our
selection. We reject M537040 (S\,Ori\,20; KJN\,52) because of its 
relatively high proper motion (Table \ref{tab_sOri_GCS:NM}), 
confirming its previously reported doubtful membership 
\citep{bejar99,barrado03a,kenyon05}. Moreover, we should
mention that M92149 has a discrepant proper motion with the mean value
of the cluster despite showing several signs of membership, including
lithium, H$\alpha$ in emission, X-ray emission \citep{franciosini06},
and a mid-infrared flux excess \citep{hernandez07}.  This object could
be part of the Orion complex and not of \sOri{}, or could have been
rejected from the $\sigma$ Orionis cluster. The remaining five 
Mayrit member candidates did not satisfy the $Z-J$ colour criterion,
including two which are borderline (Table \ref{tab_sOri_GCS:NM}),
including M487350 \citep[(SE 2004)70;][]{caballero06b} which is
discussed in more detail in Sect.\ \ref{sOri_GCS:compare_cat_other}.

We note that all of the Mayrit candidate members fainter than $J$
= 12 mag, which are confirmed spectroscopically (Table 1 of
\citet{caballero08f}), are present in our list of candidates whereas
the two objects rejected as non-spectroscopic members, Mayrit\,926051
and UCM\,0536$-$0239, are not in our catalogue. A similar result is
drawn from the sample of spectroscopic members listed in Table 1 of
\citet{sacco08}. The only exception is SE\,51, classified as a member
on the basis of radial velocity, lithium content, and H$\alpha$
emission but not present in our catalogue because it does not satisfy
the point source requirement ({\tt{Class}} parameter is not equal to
$-$2 or $-$1 as set in our SQL query). Moreover, three sources (S36,
S42, and S97), rejected by \citet{sacco08} as members, are in our
list: the first one is not in the Mayrit catalogue and the
remaining two are in the Mayrit catalogue and classified as member
candidates \citep{caballero08c}.

\subsection{The Kenyon et al.\ catalogue}
\label{sOri_GCS:compare_cat_Kenyon}

\citet{kenyon05} published a list of candidates divided into three
groups: 57 members, 13 ambiguous i.e.\ members or non-members, and 6
non-members based on their photometric ($RI$) and spectroscopic
(lithium absorption, gravity derived from the sodium doublet
equivalent width, and radial velocity) properties. All these sources
are present in the GCS database, except one member (KJN\,35)  
and one non-member (KJN\,7), because they do not satisfy our point 
source criteria (Sect.\ \ref{sOri_GCS:selection}). None of their
spectroscopic non-members are present in our list of candidates, in
agreement with their non-membership classification.  We found that the
photometry of all their spectroscopic members (Table
\ref{tab_sOri_GCS:Known_Memb}) is consistent with membership and we
verify their candidacy.  Also, we classified six out of 13
``Maybe'' candidates as photometric member candidates in our survey
(KJN\,25, 27, 40, 72, 73, and 76). KJN\,52 (also a ``Maybe'' 
object; SOri\,20) has a proper motion inconsistent with that of the 
cluster and is a binary candidate \citep{kenyon05}. The remaining six
sources (KJN\,29, 33, 34, 45, 53, and 59; Table \ref{tab_sOri_GCS:NM}) do 
not fit the cluster sequence in the colour-magnitude diagrams shown in 
Fig.\ \ref{fig_sOri_GCS:cmds_vpd}.

\subsection{The Burningham et al.\ catalogue}
\label{sOri_GCS:compare_cat_Burningham}

Using follow-up spectroscopy, \citet{burningham05a} published
absorption feature equivalent widths for 54 sources selected
photometrically by \citet{kenyon05} and radial velocities for 37 of
them.  Among these 54 sources in Table 3 of \citet{burningham05a}, 48
are in the GCS database but only 11 are within the 30 arcmin radius
from the cluster centre. All 11 passed our photometric criteria and
have membership probabilities larger than 80\%, except for one for
which equivalent width and radial velocity measurements are missing
\citep[Object 285;][]{burningham05a}. The GCS photometry is provided
in Table \ref{tab_sOri_GCS:NM}.

\subsection{The Spitzer catalogues}
\label{sOri_GCS:compare_cat_Spitzer}

\citet{hernandez07} and \citet{luhman08c} published mid-infrared
photometry obtained by the {\it{Spitzer}} telescope for candidate
members compiled from the literature, mostly based on spectroscopy
and optical and near-infrared colour-magnitude diagrams.

A total of 306 objects are common to those two catalogues. From the
sample of \citet{luhman08c}, we have found 201 sources common to our
list of candidates out of the 371 objects given in their Table 1\@.
We note that 131 of these 371 sources are brighter than $J$ = 12 mag
and therefore outside our selection criteria. The near-infrared
photometry available for the remaining 371$-$201$-$131 = 39
candidates published in Table 1 of \citet{luhman08c} suggest that
they are not probable members of \sOri{} because they lie to the blue
of the cluster sequence in at least one of the colour-magnitude
diagrams.  The Spitzer photometry published by \citet{luhman08c}
indicates that five out of 39 exhibit mid-infrared flux excesses,
including SOri\,55 which may have a disk \citep{luhman08c}. Three 
other objects, SO\,566, SO\,673, and SO\,733, show mid-infrared excesses, 
suggesting that they could be seen in scattered light or have edge-on disks, 
leading to their rejection photometrically due to unusual
colours and magnitudes.

We repeated this procedure for the sample published
in Table 1 of \citet{hernandez07}. We confirm the candidacy of
169 sources out of 336 of their probable candidates, taking into
account that another 120 are brighter than our $J$-band threshold.
Most of them are in the \citet{luhman08c} sample and included in the
Mayrit catalogue. Among the 133 uncertain members of \sOri{} listed 
in Table 2 of \citet{hernandez07}, we recovered 15 sources that 
we classify as photometric members; the remaining ones were rejected.

\subsection{Other published sources}
\label{sOri_GCS:compare_cat_other}

We also cross-correlated our list of candidates with Table 3 of
\citet{caballero07d} containing cluster member candidates fainter than the 
limiting magnitude of the Mayrit catalogue. We recovered all objects 
in the magnitude interval $J$\,=\,16.4--18.3 mag and
identified them as very likely candidates through our five-band photometry.
However, we rejected S\,Ori J053944.5$-$025959 and S\,Ori J053956.8$-$025315
from our list of very likely candidates because they appear too blue in the
($J-K$,$J$) colour-magnitude diagram.
We note that SOri J053922.2$-$024552, listed in Table 3 of
\citet{caballero07d} but not in the Mayrit catalogue, is in our list
of GCS candidates. In addition, we extracted the GCS photometry
for five candidates in Table 3 of \citet{caballero07d} but not in
our list because of the limit in $J$. These objects are
SOri J053858.6$-$025228, SOri J053949.5$-$023130, SOri\,60 (L2$\pm$1.0),
SOri J053844.5$-$025512, and SOri J054008.5$-$024550\@. We note that
the first of these sources is extremely red in $Y-J$ and has a Spitzer
detection with an upper limit in the [3.6]--[8.0] colour
\citep{caballero07d}. The last object is only detected in $JHK$.
In addition, we recovered a few additional objects originally
published by \citet{bejar99}, \citet{zapatero00}, and \citet{bejar01}.
Their photometry is also given in Table \ref{tab_sOri_GCS:Known_Memb}.

We found a few additional sources that do not fit the cluster
sequence in at least two of the colour-magnitude diagrams (crosses in
Fig.\ \ref{fig_sOri_GCS:cmds_vpd}). Those objects, rejected from
our list of photometric members and discovered by \citet{bejar99},
\citet{zapatero00}, and \citet{bejar01} are SOri\,26 (M4.5), SOri\,34,
41, 43, 44 (M7), 49 (M7.5), 57, SOri J053909$-$022814 (M5), SOri
J053926$-$022614, and SOri J053948$-$022914 (M7). Moreover, we found
that SOri\,54 (M9.5), 55 (M9), and 58 (L0) are borderline and maybe
part of an adjacent population belonging to Orion as they show
evidence of disks \citep{zapatero07b}. We consider them as
``Maybe'' candidate members but do not include them in our final
list of photometric candidate members. Table \ref{tab_sOri_GCS:NM}
lists their coordinates and GCS photometry along with the original
names.

S Ori J053948.1$-$022914 was originally identified by \citet{bejar01}
and its membership remains under debate \citep{caballero07d}. The
object has $J$ = 16.40$\pm$0.05 mag and a slightly red $J-K_{s}$
colour of 1.28$\pm$0.17 \citep{barrado03a}. However, the GCS DR4 has
$J$ = 16.382$\pm$0.013 mag and $J-K$ = 1.016$\pm$0.020 mag, the
correction from 2MASS to GCS magnitudes for an M4 dwarf being smaller
than 0.01 mag \citep{hewett06}. Both colour estimates agree within the
error bars. This object lies on the blue side of the cluster sequence
(Fig.\ \ref{fig_sOri_GCS:cmds_vpd}) but it cannot be definitively
rejected. Its membership thus remains open.

\citet{caballero06b} argued that SE\,70 (classified as M5--M6) and
SOri\,68 (L5$\pm$2 with an estimated mass of 5 M$_{\rm Jup}$) form
a wide (physical projected separation of $\sim$1700 au) system and belong 
to \sOri{} based on photometric, spectroscopic, and kinematic arguments.
The secondary, SOri\,68, is not detected on the GCS images but SE\,70
is clearly seen with $J$ = 15.276$\pm$0.006 mag. Its optical-to-infrared 
colours are $Z-J$ = 1.056 and $Y-J$ = 0.568, hence this object 
lies to the blue of the \sOri{} sequence in the ($Z-J$,$Z$) and 
($Y-J$,$Y$) diagrams (open diamond in Fig.\ \ref{fig_sOri_GCS:cmds_vpd}). 
A comparable trend was visible in the ($I-J$,$I$) diagram 
\citep{caballero06b}. Its $J-K$ colour (0.871), however, looks
consistent with the cluster sequence. The measured proper motion,
($\mu_{\alpha}\cos{\delta}$,$\mu_{\delta}$) = ($-$0.4,$-$10.3) mas/yr
(Table \ref{tab_sOri_GCS:NM}), is small and cannot be used to distinguish
between a cluster member and a very distant M dwarf. Additionally, we note 
that about 70\% of field dwarfs with a spectral type of M5 and M6 are active
\citep{west08}, suggesting that this criterion is not sufficient to 
assign a young age to SE\,70 \citep{franciosini06}.
Furthermore, the radial velocity measurement has large error bars
and the detection of the lithium absorption is noisy \citep{caballero06b}.
Therefore, we question its membership to \sOri{} and cast
doubt about the binary nature of this supposedly wide system.

Finally, based on those objects discussed in this Section that
have not been selected as likely $\sigma$ Orionis photometric members
in our GCS survey but that show signatures of a very young age
(suggesting their possible membership in the cluster), we estimate
that our search may have lost up to $\sim$8\,\% of true members of the
cluster.

\subsection{New member candidates}
\label{sOri_GCS:compare_cat_NEW}

We identified 66 new candidate members not published in previous
studies. Seven of these have additional $Iz$($JHK$) photometry 
in \citet{bejar01_PhD}. Their photometry is
indicative of cluster membership and their proper motions are not large
enough to discard them. Table \ref{tab_sOri_GCS:new_MEMB} gives
their coordinates, near-infrared photometry, and proper motions. They
are plotted as red filled circles in Fig.\
\ref{fig_sOri_GCS:cmds_vpd}.  Most of them appear to be located in the
periphery of the central 30 arcmin, possibly due to the
incompleteness of previous surveys in the outer parts of the cluster
and the (possibly) greater contamination by an older population
belonging to the Orion complex.

We cross-correlated this list of new candidates with the sample
of non-members published by \citet{caballero08c} and found two objects
in common: S Ori J053945.01$-$025348.9 and S Ori
J053929.71$-$022722.7\@.  Both sources were rejected as cluster
members because no lithium absorption is seen in good quality and
high-resolution spectra.  We checked our Spitzer database
(Zapatero Osorio et al. 2007) and found that none shows mid-infrared
flux excesses.


%

%
%
\section{Variability of low-mass stars and brown dwarfs}
\label{sOri_GCS:variability}

In this section, we examine the variability of low-mass stars and
brown dwarfs over a timescale of years by comparing the $J$-band
photometry from 2MASS \citep{cutri03} and the GCS DR4\@. We first focus
on the $J$-band because we will derive later the cluster luminosity
(Sect.\ \ref{sOri_GCS:LF}) and mass (Sect.\ \ref{sOri_GCS:MF}) 
functions from this passband. The level and intensity of the activity
would set a lower limit on the magnitude bins chosen to derive the
luminosity function.

\begin{figure}
   \centering
   \includegraphics[width=\linewidth]{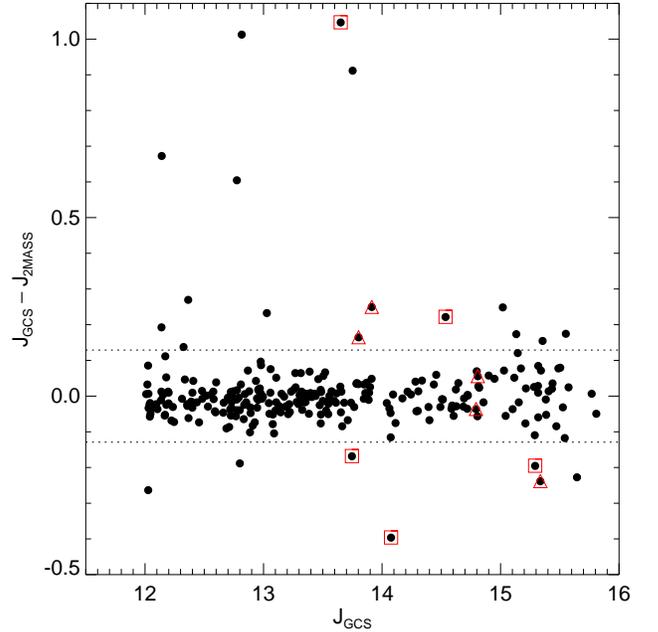}
   \caption{Difference between the $J$-band magnitudes in the
2MASS and GCS systems as a function of the $J_{\rm GCS}$ magnitude (see text).
Five objects exhibit variability over $\sim$yearly time-scales with a 99.5\%
confidence level (or 3$\sigma$; dashed lines) in $J$, $H$, and $K$ (open 
squares). Another five sources show variability in two bands (open triangles).
The other sources beyond the dashed lines show a 3$\sigma$ variability
in $J$ only.
}
   \label{fig_sOri_GCS:variability}
\end{figure}

We considered the $J$ = 12--16 magnitude range (corresponding to
masses between $\sim$0.5 and 0.03 M$_{\odot}$) where both surveys
provide reliable photometry. A total of 263 sources have 
$J$ = 12--16 mag among our GCS list of cluster member candidates.
The median difference between $J_{\rm GCS}$ and $J_{\rm 2MASS}$ 
magnitudes is $-$0.007 mag, after taking into account a mean offset of 
$\sim$0.048 mag for M dwarfs between the 2MASS and UKIDSS photometric 
systems \citep{hewett06}. The standard deviation, computed as 1.48 times 
the median absolute deviation for a robust Gaussian--equivalent root mean
square, is 0.043 mag. Thus, any object could be considered as variable (over 
a timescale of several years) with a 99.5\% confidence level if the measured 
difference in the $J$-band between both epochs is larger than 3$\sigma$.
A total of 28 objects satisfy this criterion
(see Fig.\ \ref{fig_sOri_GCS:variability}).
Two objects show differences larger than one magnitude and three others 
larger than 0.5 mag in $J$. The remaining ones have amplitudes less than 
0.3 mag, implying that the bins of 0.5 mag used to derive the luminosity 
function are adequate to avoid a significant impact on the luminosity 
function due to variability.

We repeated this procedure in $H$ and $K$ and found 13 and 24 
candidates for variability, respectively. Hence, we estimate an upper
limit of 20\% of variable sources among members of \sOri{}. Only five 
sources show variability in all three filters with amplitudes larger 
than 3$\sigma$ (open squares in Fig.\ \ref{fig_sOri_GCS:variability}; 
Table \ref{tab_sOri_GCS:GCScand_variable}), implying a minimum level of
variability of 2\%. Another two objects exhibit variability in $J$ and 
$H$ while three others show  variability in $H$ and $K$ (open triangles 
in Fig.\ \ref{fig_sOri_GCS:variability};
Table \ref{tab_sOri_GCS:GCScand_variable}). For two objects, the highest
spatial resolution and greater depth of the GCS images compared to 2MASS 
reveals a faint companion close to M726005 and M458140\@. All of the
10 candidate variables but one have $J-K >$ 1.0 mag and are likely to
harbour disks as they are clearly located to the red of the cluster
sequence in the ($J-K$,$J$) diagram (filled triangles in 
Fig.\ \ref{fig_sOri_GCS:cmds_vpd}). 
Indeed, all of them but two (M1082188 and M1583183) show signs of youth 
with H$\alpha$ in emission, presence of lithium and disk (see
Table \ref{tab_sOri_GCS:GCScand_variable} for the detections and
references).

%
%
%

%
%
\begin{figure*}
   \centering
   \includegraphics[width=\linewidth]{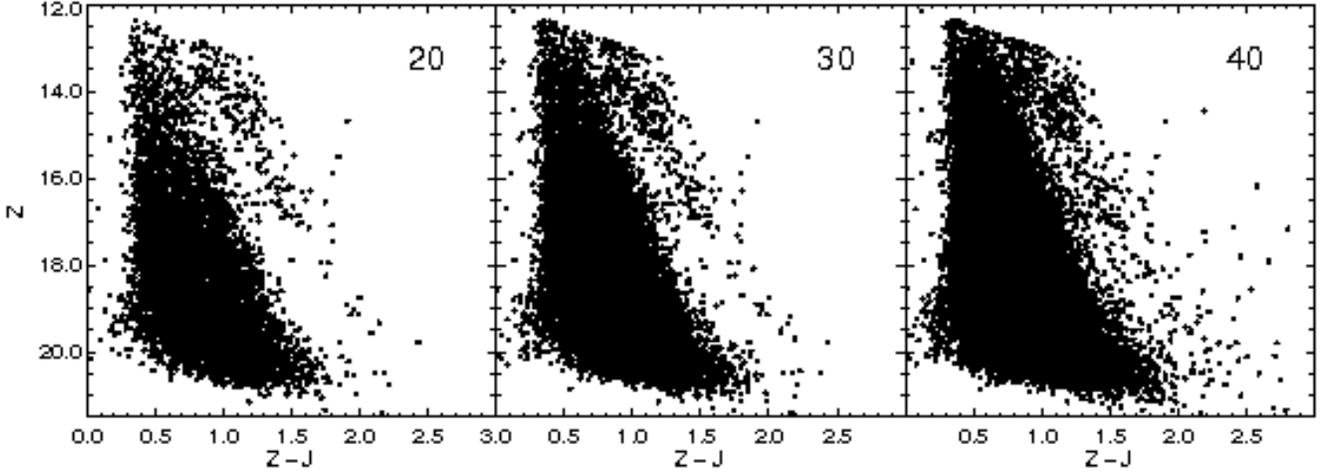}
   \caption{($Z-J$,$Z$) colour-magnitude diagrams with all detections
   within the central 20, 30, and 40 arcmin from the cluster
   centre from left to right, respectively. The cluster sequence is clearly
   distinguished from field stars up to 30 arcmin from the cluster centre.
}
   \label{fig_sOri_GCS:cmd_zjz_extent}
\end{figure*}
%

%
%
\section{The extent of the cluster}
\label{sOri_GCS:extent}

Figure \ref{fig_sOri_GCS:cmd_zjz_extent} shows ($Z-J$,$Z$)
colour-magnitude diagram drawn for concentric circles, centered on the star
$\sigma$ Ori, with radii of 20, 30, and 40 arcmin. The cluster
sequence is clearly distinguished from the field stars up to 30 arcmin
(smaller radii are not shown for clarity of the figure).
At larger radii, the contamination by the population belonging to the Orion
complex and foreground field stars increases, blurring the \sOri{}
sequence with field stars. Some cluster member candidates might, however,
be present beyond that limit. This topic is discussed in detail in
\citet{bejar04}, \citet{caballero08a}, and will be addressed in
a forthcoming publication (B\'ejar et al., in prep).

%
%
\section{The distribution of low-mass stars and brown dwarfs}
\label{sOri_GCS:distribution}

In this section, we discuss the distribution of low-mass stars and
brown dwarfs based on the member candidates extracted from the GCS\@.

Table \ref{tab_sOri_GCS:nb_VLM_BD} shows the numbers of low-mass
stars (M $\sim$ 0.5--0.08 M$_{\odot}$) and brown dwarfs (M =
0.08--0.02 M$_{\odot}$) for various distances from the cluster
centre. The largest radius used for the computation is 30 arcmin,
the limit set for our study. The GCS might be incomplete in the
inner one arcmin because of the presence of the bright star $\sigma$ Ori
itself. The masses are directly derived from the magnitudes given
by the NextGen \citep{baraffe98} and DUSTY \citep{chabrier00c} models.
Moreover, we have fixed the age to 3 Myr 
\citep{oliveira02,zapatero02a,sherry04} and considered two possible 
distances: 352 pc derived by {\it{Hipparcos}} \citep{perryman97} and 
440 pc \citep{brown94,sherry04,jeffries06}. 

The relative numbers of brown dwarfs and stars, defined as the
R$_{SS}$ ratio by \citet{briceno02} or as R1 by \citet{luhman03b}, 
is computed by adding the high-mass stars (from $\sim$10 to 0.5 M$_{\odot}$)
from the Mayrit catalogue to our list of candidates identified in 
the GCS. This ratio is roughly half the size for the greater 
distance estimate of \sOri{}. We derived R1 = 0.23$\pm$0.02 
and 0.14$\pm$0.03 for distances of 352 pc and 440 pc, respectively.
These numbers increase to 0.45$\pm$0.02 and 0.25$\pm$0.05 when we 
limit our computation to the mass range covered by the GCS 
(M$\sim$ 0.5--0.013 M$_{\odot}$).

The R1 ratios derived for Taurus \citep{luhman04a},
IC\,348 \citep{luhman03b}, and the Trapezium Cluster \citep{luhman00b}
are 0.14, 0.12, and 0.26 with an uncertainty of 0.04, respectively.
For \sOri{}, we derived 0.23$\pm$0.02 when including the highest
mass members, consistent with the values inferred for the Trapezium
Cluster. For the R2 ratio, i.e.~the ratio of high-mass (1--10 M$_{\odot}$
to low-mass (0.15--1 M$_{\odot}$) stars as defined by \citet{luhman03b}, 
we derived 0.28$\pm$0.04, again consistent with the Trapezium cluster 
\citep[0.21$\pm$0.04;][]{luhman00b} and 
also with the Pleiades \citep[0.27$\pm$0.02;][]{bouvier98}, 
M35 \citep[0.22$\pm$0.01;][]{barrado01a}, 
and Chamaeleon I \citep[0.26$\pm$0.06;][]{luhman07d} but two times
larger than for Taurus \citep[0.13$\pm$0.04;][]{luhman04a}. 
This resemblance with Trapezium is not unexpected since both it 
and \sOri{} are part of the Orion complex. 
We should emphasise the large error on the
distance of \sOri{} and the different sources of uncertainties
when comparing ratios derived in different regions, particularly
when different methods have been employed to infer masses.
Hence, we conclude that the star--brown dwarf fraction of \sOri{}
is probably similar to that of other regions.

%
%
\begin{table}
 \centering
  \caption{Comparison of the numbers of low-mass stars
and brown dwarfs as a function of the distance
from the cluster centre (D), from 5 to 30 arcmin in radius.
}
 \label{tab_sOri_GCS:nb_VLM_BD}
 \begin{tabular}{@{\hspace{0mm}}c | c c c | c c c@{\hspace{0mm}}}
 \hline
 D  & stars$^{a,b}$ & BD$^{c}$ &  R1$^{d}$ & stars$^{a}$ & BD$^{c}$ &  R1$^{e}$   \cr
    &             &     &  \%   &           &     &  \%   \cr
 \hline
 5  &  24 ( 25) &   5  & 20.8 (10.2)  &  27 ( 25) &   3  & 11.1 ( 5.8)  \cr
10  &  64 ( 43) &  24  & 37.5 (22.4)  &  77 ( 43) &  13  & 16.9 (10.8)  \cr
15  &  93 ( 62) &  37  & 39.8 (23.9)  & 109 ( 62) &  24  & 22.0 (14.0)  \cr
20  & 128 ( 81) &  51  & 39.8 (24.4)  & 147 ( 81) &  37  & 25.2 (16.2)  \cr
25  & 163 (108) &  59  & 36.2 (21.8)  & 186 (108) &  42  & 22.6 (14.3)  \cr
30  & 198 (120) &  73  & 36.9 (23.0)  & 223 (120) &  56  & 25.2 (16.4)  \cr
 \hline
 \end{tabular}
\begin{list}{}{}
\item[$^{a}$] In this table, low-mass stars are defined as objects with masses 
between 0.5 and 0.08 M$_{\odot}$ (for the purpose of the indices)
\item[$^{b}$] The numbers in brackets take into account the most massive members 
($J \leq$ 12 mag; M $\geq$ 0.5 M$_{\odot}$) of \sOri{} taken Mayrit catalogue 
\citep{caballero08c}, yielding the R$_{SS}$ or R1 ratios defined by \citet{briceno02} 
and \citet{luhman03b}, respectively
\item[$^{c}$] In this table, brown dwarfs are defined as objects with masses 
between 0.08 and 0.02 M$_{\odot}$ (for the purpose of the indices)
\item[$^{d}$] The results on the left-hand side assume a distance of 352 pc from {\it{Hipparcos}}
\item[$^{e}$] The results on the right-hand side assume a distance of 440 pc \citep{brown94,sherry04}
\end{list}
\end{table}
%

%
%
\begin{table}
 \centering
  \caption{Cumulative numbers of stars (M = 0.33--0.075 M$_{\odot}$) 
  and brown dwarfs (M = 0.075--0.013 M$_{\odot}$) as a function of radius
  (in arcmin) from the cluster center.
}
 \label{tab_sOri_GCS:number_star_BD_radius}
 \begin{tabular}{l c c c c c c}
 \hline
radius &   5  &  10 &  15 &  20 &  25 &  30 \cr
 \hline
stars  &  24  &  65 &  94 & 129 & 165 & 201 \\
BDs    &   6  &  27 &  41 &  57 &  67 &  82 \\
 \hline
 \end{tabular}
\end{table}
%

%
%
\begin{figure}
   \centering
   \includegraphics[width=\linewidth]{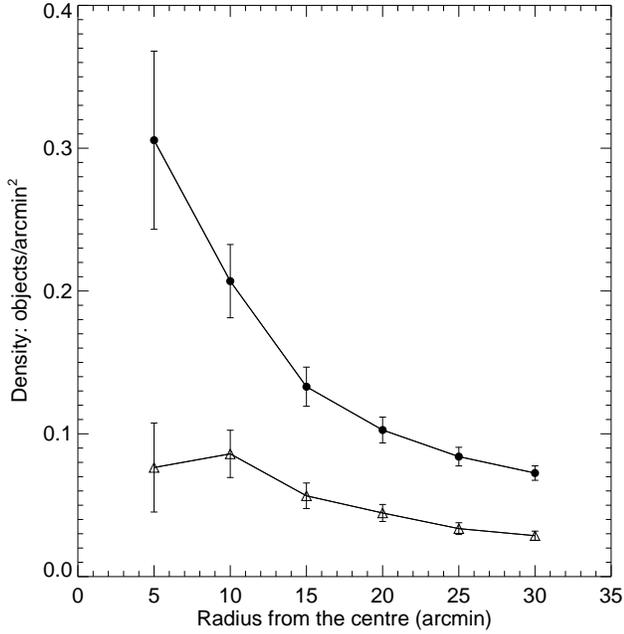}
   \caption{Histogram of the numbers of stars (filled circles)
   and brown dwarfs (open triangles) per square arcmin as a 
   function of radius (in arcmin) from the cluster center
   (Table \ref{tab_sOri_GCS:number_star_BD_radius}).
   Poisson errors are assumed.
}
   \label{fig_sOri_GCS:histo_star_BD}
\end{figure}

The R1 ratios appear fairly constant as a function of radius although
we observe a possible dearth of brown dwarfs within the central five
arcmin (Table \ref{tab_sOri_GCS:nb_VLM_BD}; Fig.\
\ref{fig_sOri_GCS:histo_star_BD}) even after correcting by a factor of
0.04 (1$^{2}$/5$^{2}$) for the saturation within the central one
arcmin generated by the bright star $\sigma$ Ori \citep[$J$ =
4.75;][]{cutri03}.  Table \ref{tab_sOri_GCS:number_star_BD_radius}
gives the numbers of stars and brown dwarfs as a function of radius
from the cluster. This distribution is illustrated in Fig.\
\ref{fig_sOri_GCS:histo_star_BD}. The number of brown dwarfs seems to
drop off or flatten in the central five arcmin whereas the number of
stars keeps rising.

This dearth of brown dwarfs remains independently of the choice of the 
cluster's distance (Table \ref{tab_sOri_GCS:nb_VLM_BD}). This result, 
already noticed by \citet{caballero08a}, 
could be explained by the presence of $\sigma$ Ori, a massive multiple 
system at the center of the cluster \citep{caballero08b} that would have 
prevented the survival of the lowest-mass objects in the core of 
the cluster. However, new potential member candidates were extracted 
within the central 1.5$\times$1.5 arcmin of the cluster from
a multi-conjugate adaptive optics survey \citep{bouy09a}, suggesting
that this issue is not yet settled.
The formation mechanism of brown dwarfs proposed by \citet{whitworth04}, 
where the wind of O and B stars would erode the cores during the formation, 
would not be the main mechanism at play in \sOri{}.
If we assume no dynamical evolution, then the dearth of brown dwarfs
tells us about the formation of stars and brown dwarfs, namely that
low-mass objects are not found very close to the center but beyond
the core radius \citep[about six arcmin;][]{kharchenko05a}.
If we assume, to the contrary, that the cluster has suffered dynamical
evolution (even at 3 Myr), then some fraction of the lowest mass stars and brown
dwarfs would have been ejected beyond the cluster radius 
\citep[about 24 arcmin;][]{kharchenko05a}. Our data allows us to set 
a lower limit of 30 arcmin as we do not see an excess of low-mass objects 
within the central 30 arcmin (Fig.\ \ref{fig_sOri_GCS:histo_star_BD}).

%
%
\begin{figure}
   \centering
   \includegraphics[width=\linewidth]{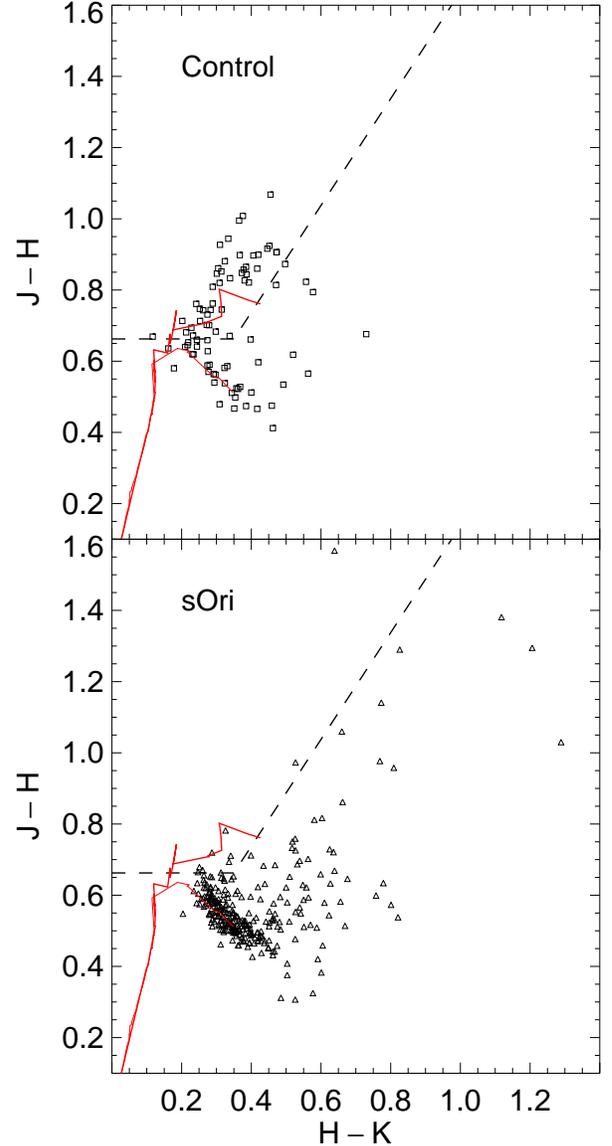}
   \caption{($H-K$,$J-H$) two-colour diagrams for the candidates
identified in the \sOri{} (upper panel) and the control field (lower panel).
The red lines represent the stellar main-sequence and the giant branch
\citep{hewett06}.
}
   \label{fig_sOri_GCS:ccd_HKJH_contam}
\end{figure}
%

%
%
%
\section{Level of contamination}
\label{sOri_GCS:contamination}

We have attempted to address the issue of photometric contamination
by applying our selection procedure to one control field with the same
galactic latitude as \sOri{}. This field is located in the Orion complex
but far enough from \sOri{} and the Trapezium Cluster to minimize the
effect of extinction and the presence of young objects
(see Fig.\ \ref{fig_sOri_GCS:coverage}). However, because
this control field is located in the Orion complex, the levels of
contamination estimated in this section are very likely upper limits.

We selected the area delineated by RA=84.5--85.5 deg and Dec 
between $-$4.78 and $-$4.0 deg i.e. south--east of \sOri{}; this area 
was chosen to match the coverage in \sOri{} under study.
We repeated the photometric and proper motion selection steps described
in Sect.\ \ref{sOri_GCS:selection_CMDs} \& \ref{sOri_GCS:selection_PM}.
The sources extracted in the control field after applying all
selection criteria are plotted in the ($H-K$,$J-H$) two-colour diagram
shown in the upper panel of Fig.\ \ref{fig_sOri_GCS:ccd_HKJH_contam}.
The total number of sources is 87\@. \citet{lucas08} show ($H-K$,$J-H$)
two-colour diagrams at different galactic latitudes from the UKIDSS 
Galactic Plane Survey to identify the multiple groups of stars present
along the line of sight. By comparison with our diagram, we can
immediately conclude that a large number of contaminants identified
in the control field are giants (Fig.\ \ref{fig_sOri_GCS:ccd_HKJH_contam}).
For comparison, the same diagram is shown in the lower panel for all 
candidates identified in \sOri{}. Therefore, to estimate the true level 
of contamination in our cluster sample, we have kept only sources below
the dashed lines in Fig.\ \ref{fig_sOri_GCS:ccd_HKJH_contam} defined by:
\begin{itemize}
\item[$\bullet$] $H-K \leq$ 0.35 and $J-H \leq$ 0.6625
\item[$\bullet$] $J-H \leq$ 1.5$\times$($H-K$) $+$ 0.1375
\end{itemize}

The numbers of contaminants (dN$_{\rm CF}$) in each interval of magnitude 
are given in Table \ref{tab_sOri_GCS:LF_MF_numbers}. They are typically 
within the Poisson errors of the numbers of candidate members identified 
in \sOri{}. The level of contamination is around 15\% for low-mass stars,
13\% for brown dwarfs, and rises to $\sim$35\% for very low-mass brown
dwarfs and planetary--mass objects with masses below 0.02 M$_{\odot}$
(Table \ref{tab_sOri_GCS:LF_MF_numbers}).

%
%
\section{The luminosity and mass functions}
\label{sOri_GCS:MF_LF}

In this section, we derive the cluster (system) luminosity and mass 
functions based on 285 cluster member candidates confirmed photometically
and the additional nine sources selected in the ($Y-J$,$Y$) and ($J-K$,$J$)
diagrams (dN in Table \ref{tab_sOri_GCS:LF_MF_numbers}). We have subtracted 
from this list the potential contaminants identified in the control field
(dN$_{\rm CF}$ in Table \ref{tab_sOri_GCS:LF_MF_numbers}; total 47 sources).
Hence, the luminosity and mass functions are derived using the difference of 
both lists (dN$_{\rm corr}$ in Table \ref{tab_sOri_GCS:LF_MF_numbers}).
No correction for binaries is applied to the luminosity and mass functions.
Finally, we put our results into context and compare them with previous 
studies in \sOri{} and other clusters.

For the rest of the paper, we will assume an age of 3 Myr and a distance 
of 352 pc for \sOri{} unless stated otherwise. While these parameters have 
been used extensively in the literature as the most probable values, a
word of caution is necessary before discussing the shape of the
luminosity and mass functions. \citet{jeffries06} demonstrated
that two populations (see also \citet{caballero07b}), with distinct 
ages (3 vs 10 Myr) and distances (352 vs 440 pc), are indistinguishable 
in colour-magnitude diagrams. However, the contamination is less
than 10\% within the central 10 arcmin but can go up to 40--50\%
towards the north--west of \sOri{} due to the presence of another
bright star, $\zeta$ Ori.
Therefore, any list of candidate members belonging to \sOri{} will
be a mix of two populations, leading to some level of errors in the
mass determination \citep[see extensive discussion in][]{jeffries06}. 
The main implication is that the ``true'' mass 
function of \sOri{} can only be derived once accurate radial velocity
measurements are available for all candidates. Keeping these caveats
in mind, we now proceed with the analysis of the luminosity 
and mass functions.

\subsection{The luminosity function}
\label{sOri_GCS:LF}

The final sample contains 285$+$9 = 294 photometric member candidates in 
the $J$ = 12--19 mag range within the central 30 arcmin from the cluster 
centre. We have removed 47 objects identified as contaminants in the
control field, leaving 294$-$47 = 247 sources. If we assume an age of 
3 Myr and a distance of 352 pc,
models \citep{baraffe98,chabrier00c} predict that this sample contains
169 stars and 82 brown dwarfs. However, if we assume the larger
distance, the numbers of stars and brown dwarfs would be 195 and 56,
respectively.

\begin{figure}[!h]
   \centering
   \includegraphics[width=\linewidth]{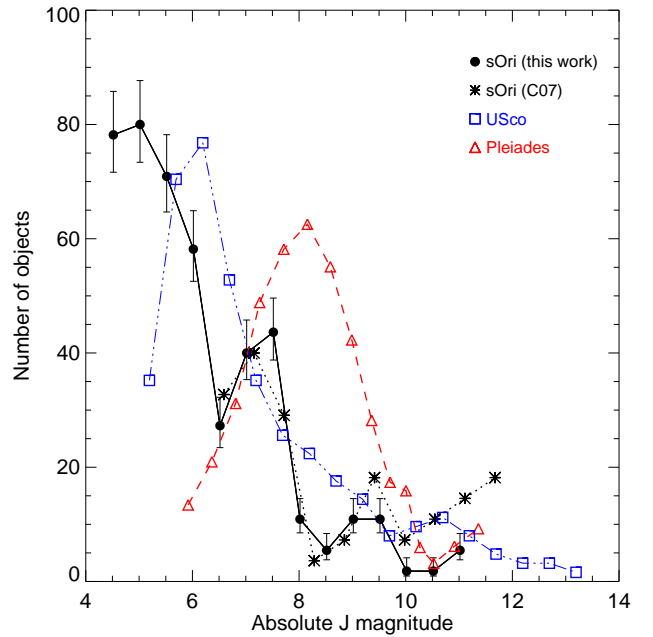}
   \caption{The luminosity function for \sOri{} (filled circles; see text).
Overplotted are the
luminosity functions for sOri \citep[star symbols;][]{caballero07d},
Upper Sco \citep[open squares; 5 Myr, 145 pc;][]{lodieu07a}, and
the Pleiades \citep[open triangles; 125 Myr, 130 pc;][]{lodieu07c}.
}
   \label{fig_sOri_GCS:LFplot}
\end{figure}

Figure \ref{fig_sOri_GCS:LFplot} displays the luminosity function
i.e.\ the numbers of stars as a function of absolute $J$ magnitude (per 
0.5 mag bins) for \sOri{} and two other clusters targeted by the GCS in a 
similar fashion: Upper Sco \citep[5 Myr, 145 pc;][]{lodieu07a} and
the Pleiades \citep[125 Myr, 130 pc;][]{lodieu07c}.
They are all normalised to \sOri{} at $J \sim$ 14.75 mag, corresponding
to M$_{J} \sim$ 7 mag.
Error bars are Gehrels errors \citep{gehrels86} rather than Poissonian 
error bars because the former represent a better approximation to the
true error for small numbers. The upper limit is defined as
1+($\sqrt(dN+0.75)$) and the lower limit as $\sqrt(dN-0.25)$
assuming a confidence level of one sigma.

Three peaks are visible in the luminosity function of \sOri{} at 
$J \sim$ 5.25, 7.25, and 9.25 mag. The latter two peaks are also
seen in the luminosity function presented in \citet{caballero07d},
shown as star symbols in Fig.\ \ref{fig_sOri_GCS:LFplot}.
In the other clusters, we observe two peaks with a dip around
M$_{J}$ = 9.5 and 10.5 mag in Upper Sco and the Pleiades, respectively, 
likely due to the M7--M8 gap \citep{dobbie02b}.
We should emphasize that the luminosity function of Upper Sco is
``smoothed'' as we are counting the number of objects per magnitude
bin by steps of 0.5 mag \citep{lodieu07a}.
The difference in the number of peaks could arise from the different 
mass intervals probed due to the variety of ages, the extremes being 
120 Myr for the Pleiades and 3 Myr for \sOri{}.

To address the role of the older population on the determination
of the luminosity and mass functions, we have derived them
excluding all the sources lying beyond a radius of 10 arcmin towards
the north--west of the cluster where the contamination by the older
group is the largest \citep[see Fig.\ 2 of][]{jeffries06}. The numbers 
of sources as a function of magnitude (and thus mass) are given as
dN$_{\rm NW}$ in the sixth column of 
Table \ref{tab_sOri_GCS:LF_MF_numbers}.
Typically, these numbers are upper limits as we are assuming 100\% 
contamination by the old population in this region and are smaller 
than the error bars (assuming either Poisson or Gehrels errors) and 
do not have a strong influence on the luminosity (and therefore mass) 
function(s). The only word of caution concerns the $J$ = 15.0--15.5 
magnitude range, implying that the second peak seen in the luminosity 
function could be the most affected (and thus overestimated) by this
older group comparatively to the rest.

%
%
%

%
%
\begin{table*}
 \centering
  \caption{Numbers for the luminosity and mass functions
}
 \label{tab_sOri_GCS:LF_MF_numbers}
 \begin{tabular}{c c c c c c c c c}
 \hline
Jmag range   &  Mass range$^{a}$ & dM$^{b}$  &  dN$^{c}$  & dN$_{\rm CF}^{d}$ &
dN$_{\rm NW}^{e}$ & dN$_{\rm corr}^{f}$ & dN$^{g}_{bol_{\rm corr}}$  & dN/dM \cr
 \hline
12.00--12.50 &  0.4910-0.3440 &  0.1470 & 44.0 &    1.0        &     3.0      &    43.0 &  36   &  292.52  \cr 
12.50--13.00 &  0.3440-0.2460 &  0.0980 & 55.0 &   11.0        &     5.0      &    44.0 &  35   &  448.98  \cr 
13.00--13.50 &  0.2460-0.1690 &  0.0770 & 47.0 &    8.0        &     3.0      &    39.0 &  50   &  506.49  \cr
13.50--14.00 &  0.1690-0.1100 &  0.0590 & 37.0 &    5.0        &     3.0      &    32.0 &  38   &  542.37  \cr
14.00--14.50 &  0.1100-0.0670 &  0.0430 & 20.0 &    5.0        &     1.0      &    15.0 &  16   &  348.84  \cr
14.50--15.00 &  0.0670-0.0470 &  0.0200 & 23.0 &    1.0        &     2.0      &    22.0 &  23   & 1100.00  \cr
15.00--15.50 &  0.0470-0.0360 &  0.0110 & 27.0 &    3.0        &     6.0      &    24.0 &  28   & 2181.82  \cr
15.50--16.00 &  0.0360-0.0285 &  0.0075 &  8.0 &    2.0        &     1.0      &     6.0 &   1   &  800.00  \cr
16.00--16.50 &  0.0285-0.0233 &  0.0052 &  5.0 &    2.0        &     0.0      &     3.0 &   5   &  576.92  \cr
16.50--17.00 &  0.0233-0.0182 &  0.0051 &  6.0 &    0.0        &     0.0      &     6.0 &   8   & 1176.47  \cr
17.00--17.50 &  0.0182-0.0141 &  0.0041 &  8.0 &    2.0        &     0.0      &     6.0 &   3   & 1463.41  \cr
17.50--18.00 &  0.0141-0.0114 &  0.0027 &  2.0 &    1.0        &     0.0      &     1.0 &   1   &  370.37  \cr
18.00--18.50 &  0.0114-0.0092 &  0.0022 &  3.0 &    2.0        &     0.0      &     1.0 &   2   &  454.55  \cr
18.50--19.00 &  0.0092-0.0075 &  0.0017 &  7.0 &    4.0        &     0.0      &     3.0 &   3   &  588.24  \cr
 \hline
 \end{tabular}
\begin{list}{}{}
\item[$^{a}$] Mass range associated to the magnitude range given in the first column: we assumed an age of 3 Myr and a distance of 352 pc
\item[$^{b}$] Difference between the masses, dM
\item[$^{c}$] Numbers of candidate members identified in \sOri{}
\item[$^{d}$] Numbers of contaminants identified in the control field (CF)
\item[$^{e}$] Numbers of objects located beyond 10 arcmin towards the North-West of the cluster center, where the contamination by the old population is the highest
\item[$^{f}$] Numbers of candidate members in \sOri{} using method \#1 and corrected for contamination
\item[$^{g}$] Numbers of candidate members in \sOri{} using method \#2 after correction for contamination estimated from the control field
\item[$^{h}$] Values for the mass function in linear scale (dN/dM) using method \#1
\end{list}
\end{table*}

\subsection{The mass function}
\label{sOri_GCS:MF}

The direct conversion from observed magnitudes to masses relies on the
3 Myr NextGen \citep{baraffe98} and DUSTY \citep{chabrier00c}
isochrones shifted to the distance of the cluster. A larger distance
of 440 pc will lead to larger masses but should not affect notably the
shape of the mass function. This method is referred to as method \#1 
in Fig.\ \ref{fig_sOri_GCS:MFplot_sOri}. The estimated masses, spanning
$\sim$0.5 M$_{\odot}$ to $\sim$0.01 M$_{\odot}$, are shown on the
right-hand side of the colour-magnitude diagrams in Fig.\
\ref{fig_sOri_GCS:cmds_vpd}. The corresponding mass
function (in linear scale; dN/dM) is displayed with black dots in
Fig.\ \ref{fig_sOri_GCS:MFplot_sOri} with numbers in the last column
of Table \ref{tab_sOri_GCS:LF_MF_numbers}.

However, the magnitude predictions of the models fail to neatly
reproduce the observed infrared colours (specifically $J-K$) of the
coolest candidates with spectral types of late-M and
early-L\@. Consequently, our mass estimates are possibly
underestimated. In the next section (Sect.\
\ref{sOri_GCS:MF_compare}), we will employ an alternative method
(method \#2) less prone to large errors in model magnitudes to convert
apparent magnitudes into masses.


%
\subsection{Comparison with previous studies in \sOri{}}
\label{sOri_GCS:MF_compare}

\begin{figure}
   \centering
   \includegraphics[width=\linewidth]{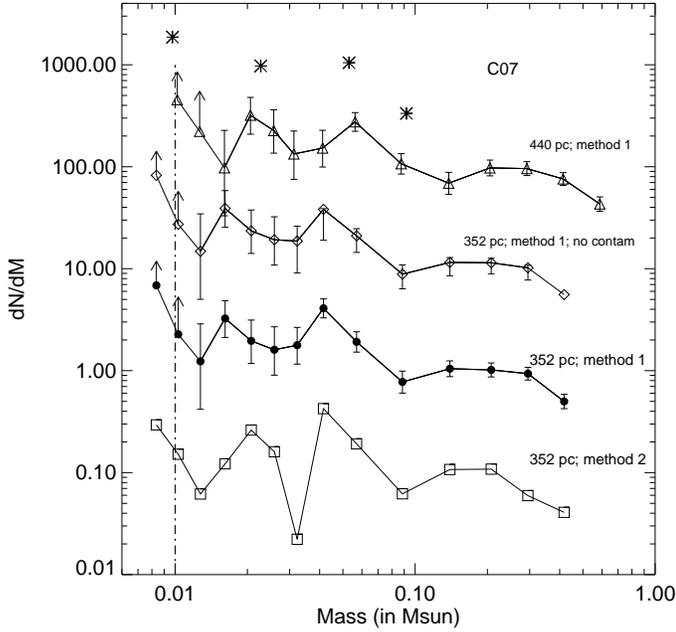}
   \caption{The mass function for \sOri{} assuming an age of 3 Myr and a 
distance of 352 pc (filled circles; Table \ref{tab_sOri_GCS:LF_MF_numbers}). 
The conversion from magnitudes into masses 
using method \#1 (filled dots) and method \#2 (open squares) is described in the
text. The mass function assuming a greater distance of 440 pc (open triangle)
is also plotted as well as the one excluding contaminants from the older
population (open diamonds; Sect.\ \ref{sOri_GCS:LF}).
A previous estimate from \citet{caballero07d} is added for comparison:
horizontal lines represent the mass range for each point.
The error bars and lower limits (arrows) are shown. The mass 
functions are shifted along the y-axis for clarity.
The vertical dot-dashed line represents the completeness of the
survey in \sOri{}.
}
   \label{fig_sOri_GCS:MFplot_sOri}
\end{figure}

The most recent estimate of the slope of the mass function in the low-mass
star, substellar, and planetary-mass regimes suggests a power law index
$\alpha$ (defined as dN/dM$\propto$M$^{-\alpha}$) equal to 0.6$\pm$0.2
in the 0.11--0.006 M$_{\odot}$ mass range
\citep[star symbols in Fig.\ \ref{fig_sOri_GCS:MFplot_sOri};][]{caballero07d}
from a survey of 0.27 square degrees in \sOri{}. This value is an
extension of the 0.23 square degree optical/infrared survey by \citet{bejar01}.
Moreover, \citet{gonzales_garcia06} derived an index $\alpha$ of
0.6$^{+0.5}_{-0.1}$ over 0.072--0.007 M$_{\odot}$ from a deep $i$,$z$
survey of 0.3 square degrees. All these estimates agree within the error
bars and probe roughly the same mass interval (see also Bihain et al.\ 2009,
subm.\ to A\&A).

To compare directly our mass function for \sOri{} with previous estimates,
we have converted the mass and luminosity parameters provided by the
NextGen and DUSTY models into observables (method \# 2; dN$_{bol_{corr}}$
in Table \ref{tab_sOri_GCS:LF_MF_numbers}) i.e., $J-K$ colours
and $J$-band magnitudes after applying bolometric corrections available from
the literature for field M and L dwarfs \citep{dahn02,golimowski04a}.
This procedure is also subject to errors since we are dealing with corrections
valid for old field M and L dwarfs, whose colours differ slightly from
younger objects \citep{jameson08b}. However, no bolometric corrections are
currently available for M and L dwarfs at ages younger than $\sim$1\,Gyr.
The resulting mass function using method \#2 is shown with open squares
for 352 pc in Fig.\ \ref{fig_sOri_GCS:MFplot_sOri} and is very similar
to the mass function derived with method \#1 (filled circles).
Mass estimates should, however, be more conservative. The power law
index obtained for our mass function, in the
dN/dM$\propto$M$^{-\alpha}$ form, is $\alpha$ = 0.5$\pm$0.2 over the
0.49--0.01 M$_{\odot}$ mass range (Fig.\ \ref{fig_sOri_GCS:MFplot_sOri}).
Our result is in agreement with published mass functions in \sOri{} but
has a wider mass range and a better statistics due to the larger area
surveyed (0.78 square degrees).

For completeness, we also display the mass function derived with method \#1
excluding the sources located beyond 10 arcmin towards the north--west
(Fig.\ \ref{fig_sOri_GCS:MFplot_sOri}; Sect.\ \ref{sOri_GCS:LF}).
This correction smoothes the peak observed
around 40 Jupiter masses but does not affect the shape of the mass
function (and thus the power law index derived above).

%
%
%
%

%
%
%
\begin{figure}
   \centering
   \includegraphics[width=\linewidth]{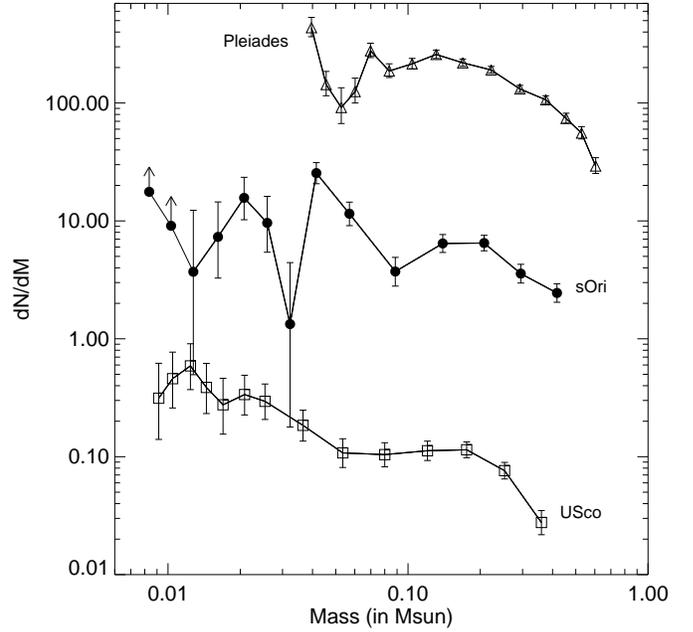}
   \caption{The mass function for \sOri{} derived with method \#2, 
assuming an age of 3 Myr and a distance of 352 pc in linear units 
(dN/dM$\propto$M$^{-\alpha}$). Also plotted are the mass functions for 
Upper Sco
the Pleiades (see text).
The error bars and lower limits (arrows) are shown. The mass functions 
are shifted along the y-axis for clarity.
}
   \label{fig_sOri_GCS:MFplot_compare}
\end{figure}
\subsection{Comparison with other clusters}
\label{sOri_GCS:other_cluster}

Fig.\ \ref{fig_sOri_GCS:MFplot_compare} compares the mass function of
\sOri{} (filled symbols) obtained with method \#2 with three regions 
with different ages and distances observed in an homogeneous manner by 
the GCS: Upper Sco \citep[open squares; 5 Myr; 145 pc;][]{lodieu07a}, and
the Pleiades \citep[open triangles; 125 Myr; 130 pc;][]{lodieu07c}.
Error bars are Gehrels error bars \citep{gehrels86}.

The \sOri{} mass function is rising towards lower masses (in a 
linear scale) and shows a strong dip at 0.03 M$_{\odot}$. The Pleiades
mass function presents a similar shape and also possess a dip at
0.05 M$_{\odot}$, attributed to the M7/M8 gap and the change in the 
properties of the atmospheres of brown dwarfs \citep{dobbie02b} and
in particular the onset of dust condensation. The index of the power 
law fit to the \sOri{} and Upper Sco mass functions is comparable within 
the error bars  over the same mass range, from low-mass stars down to the 
deuterium-burning limit. A dip is seen in the Upper Sco colour-magnitude
diagrams around $Z$ = 15.5 mag ($J \sim$ 14 mag; M $\sim$ 0.03 M$_{\odot}$)
but does not appear in the mass function due to the smoothing of the
luminosity function (dictated by the small number statistics as we
surveyed a small percentage of the full association).
These results are also in agreement with the extrapolation of the field
mass function in a log-normal form \citep{kroupa02,chabrier03}.
The mass function in \sOri{} requires spectroscopic follow-up for
the new candidates identified in this paper while the Upper Sco mass
function is quite reliable below 0.03 M$_{\odot}$ as all sources have
been confirmed spectroscopically as members \citep{lodieu08a}.

%
%
\section{Conclusions and outlook}
\label{sOri_GCS:conclusions}

We have presented the analysis of a 0.78 square degree area observed by 
the UKIDSS Galactic Clusters Survey in the young (3 Myr) and nearby
(d = 352 pc) \sOri{} cluster. The main results of our study can be 
summarised as follows:

\begin{enumerate}
\item We have confirmed photometrically the candidacy of the 
large majority of previous members identified in optical and near-infrared 
surveys. Additionally, we provide proper motions for the brightest
members, down to approximately 30 Jupiter masses. Adding the brightest
members ($J \leq$ 12 mag) of the Mayrit catalogue to our catalogue,
the population of \sOri{} is now complete from high-mass stars
down to the deuterium-burning limit within the central 30 arcmin
of the cluster
\item The distribution of low-mass stars and brown dwarfs
seems uniform from five to 30 arcmin from the cluster centre.
We find that the ratio of brown dwarfs to low-mass stars is however
smaller within the central five arcmin, possibly due to the presence
of the massive $\sigma$ Ori star (crowding and photometric incompleteness
is only affecting the central 30 arcsec from the central O star)
\item The cluster extends all the way to 30 arcmin from the centre.
Beyond this limit, the cluster sequence becomes confused with the
older members of the Orion complex and field stars
\item Five objects among the 261 low-mass stars and brown dwarfs more 
massive than 30 Jupiter masses exhibit variability over a timescale 
of years with a 95\% confidence level in $JHK$
\item The luminosity function shows three peaks at $J \sim$ 13, 15,
and 17 mag (d = 352 pc), consistent with previous studies
\item The cluster mass function shows a power--law index $\alpha$ = 0.5$\pm$0.2
in the 0.5--0.01 M$_{\odot}$ mass range in agreement with previous estimates,
and matches the slope of the Upper Sco mass function over the same mass range
\end{enumerate}

Our study provides a census of low-mass stars and brown dwarfs 
(0.5--0.01 M$_{\odot}$)
in \sOri{} and represents a reference point for deeper studies dedicated
to the search for T dwarfs in the cluster (see Bihain et al.\ 2009,
subm.\ to A\&A). The advent of the Visible and 
Infrared Survey Telescope for Astronomy (VISTA)\footnote{More details
can be found at: http://www.vista.ac.uk/} would certainly allow the
study of the fragmentation limit and the discovery of the lowest mass
entities that star formation processes can create \citep{low76,boss01} ---
the discovery and confirmation of young T dwarfs in \sOri{} is now possible.

%
%
\begin{acknowledgements}
This work was partially funded by the Ram\'on y Cajal fellowship
number 08-303-01-02\@. Financial support from the Spanish Ministry
of Science through grant AYA2007-67458 is acknowledged.
We are grateful to Isabelle Baraffe for providing us with the
NextGen and DUSTY models for the UKIRT/WFCAM filters.
We thank Jos\'e Caballero and V\'ictor S\'anchez B\'ejar for their 
comments and suggestions on the original version of the manuscript.
We thank the referee, Kevin Luhman, for his constructive report that
significantly improved the paper.

The United Kingdom Infrared Telescope is operated by the Joint
Astronomy Centre on behalf of the U.K.\ Science Technology and
Facility Council.

This research has made use of the Simbad and Vizier databases, operated 
at the Centre de Donn\'ees Astronomiques de Strasbourg (CDS), and
of NASA's Astrophysics Data System Bibliographic Services (ADS).

This publication makes use of data products from the Two Micron 
All Sky Survey (2MASS), which is a joint project of the University 
of Massachusetts and the Infrared Processing and Analysis 
Center/California Institute of Technology, funded by the National 
Aeronautics and Space Administration and the National Science Foundation.

\end{acknowledgements}

%
%
  \bibliographystyle{aa}
  \bibliography{../mnemonic,../biblio_old}

\begin{thebibliography}{78}
\expandafter\ifx\csname natexlab\endcsname\relax\def\natexlab#1{#1}\fi

\bibitem[{{Baraffe} {et~al.}(1998){Baraffe}, {Chabrier}, {Allard}, \&
  {Hauschildt}}]{baraffe98}
{Baraffe}, I., {Chabrier}, G., {Allard}, F., \& {Hauschildt}, P.~H. 1998, A\&A,
  337, 403

\bibitem[{{Barrado y Navascu{\' e}s} {et~al.}(2003){Barrado y Navascu{\' e}s},
  {B{\' e}jar}, {Mundt}, {Mart{\'{\i}}n}, {Rebolo}, {Zapatero Osorio}, \&
  {Bailer-Jones}}]{barrado03a}
{Barrado y Navascu{\' e}s}, D., {B{\' e}jar}, V.~J.~S., {Mundt}, R., {et~al.}
  2003, A\&A, 404, 171

\bibitem[{{Barrado y Navascu{\' e}s} {et~al.}(2001){Barrado y Navascu{\' e}s},
  {Zapatero Osorio}, {B{\' e}jar}, {Rebolo}, {Mart{\'{\i}}n}, {Mundt}, \&
  {Bailer-Jones}}]{barrado01c}
{Barrado y Navascu{\' e}s}, D., {Zapatero Osorio}, M.~R., {B{\' e}jar},
  V.~J.~S., {et~al.} 2001, A\&A, 377, L9

\bibitem[{{Barrado y Navascu{\'e}s} {et~al.}(2001){Barrado y Navascu{\'e}s},
  {Stauffer}, {Bouvier}, \& {Mart{\'{\i}}n}}]{barrado01a}
{Barrado y Navascu{\'e}s}, D., {Stauffer}, J.~R., {Bouvier}, J., \&
  {Mart{\'{\i}}n}, E.~L. 2001, ApJ, 546, 1006

\bibitem[{{B\'ejar}(2001)}]{bejar01_PhD}
{B\'ejar}, V. J.~S. 2001, PhD thesis, University of La Laguna, Tenerife, Spain

\bibitem[{{B{\'e}jar} {et~al.}(2004){B{\'e}jar}, {Caballero}, {Rebolo},
  {Zapatero Osorio}, \& {Barrado y Navascu{\'e}s}}]{bejar04}
{B{\'e}jar}, V.~J.~S., {Caballero}, J.~A., {Rebolo}, R., {Zapatero Osorio},
  M.~R., \& {Barrado y Navascu{\'e}s}, D. 2004, Astrophys. Space. Sci., 292,
  339

\bibitem[{{B{\'e}jar} {et~al.}(2001{\natexlab{a}}){B{\'e}jar}, {Mart{\'{\i}}n},
  {Zapatero Osorio}, {Rebolo}, {Barrado y Navascu{\' e}s}, {Bailer-Jones},
  {Mundt}, {Baraffe}, {Chabrier}, \& {Allard}}]{bejar01}
{B{\'e}jar}, V.~J.~S., {Mart{\'{\i}}n}, E.~L., {Zapatero Osorio}, M.~R.,
  {et~al.} 2001{\natexlab{a}}, ApJ, 556, 830

\bibitem[{{B{\'e}jar} {et~al.}(1999){B{\'e}jar}, {Zapatero Osorio}, \&
  {Rebolo}}]{bejar99}
{B{\'e}jar}, V.~J.~S., {Zapatero Osorio}, M.~R., \& {Rebolo}, R. 1999, ApJ,
  521, 671

\bibitem[{{B{\'e}jar} {et~al.}(2001{\natexlab{b}}){B{\'e}jar}, {Zapatero
  Osorio}, {Rebolo}, {Barrado Y Navascu{\'u}es}, {Bailer-Jones}, \&
  {Mundt}}]{bejar01a}
{B{\'e}jar}, V.~J.~S., {Zapatero Osorio}, M.~R., {Rebolo}, R., {et~al.}
  2001{\natexlab{b}}, in Astronomical Society of the Pacific Conference Series,
  Vol. 223, 11th Cambridge Workshop on Cool Stars, Stellar Systems and the Sun,
  ed. R.~J. {Garcia Lopez}, R.~{Rebolo}, \& M.~R. {Zapaterio Osorio}, 1519

\bibitem[{{Boss}(2001)}]{boss01}
{Boss}, A.~P. 2001, ApJL, 551, L167

\bibitem[{{Bouvier} {et~al.}(1998){Bouvier}, {Stauffer}, {Mart\'{\i}n},
  {Barrado y Navascu\'es}, {Wallace}, \& {B\'ejar}}]{bouvier98}
{Bouvier}, J., {Stauffer}, J.~R., {Mart\'{\i}n}, E.~L., {et~al.} 1998, A\&A,
  336, 490

\bibitem[{{Bouy} {et~al.}(2009){Bouy}, {Hu{\'e}lamo}, {Mart{\'{\i}}n},
  {Marchis}, {Barrado Y Navascu{\'e}s}, {Kolb}, {Marchetti}, {Petr-Gotzens},
  {Sterzik}, {Ivanov}, {K{\"o}hler}, \& {N{\"u}rnberger}}]{bouy09a}
{Bouy}, H., {Hu{\'e}lamo}, N., {Mart{\'{\i}}n}, E.~L., {et~al.} 2009, A\&A,
  493, 931

\bibitem[{{Brice{\~ n}o} {et~al.}(2002){Brice{\~ n}o}, {Luhman}, {Hartmann},
  {Stauffer}, \& {Kirkpatrick}}]{briceno02}
{Brice{\~ n}o}, C., {Luhman}, K.~L., {Hartmann}, L., {Stauffer}, J.~R., \&
  {Kirkpatrick}, J.~D. 2002, ApJ, 580, 317

\bibitem[{{Brown} {et~al.}(1994){Brown}, {de Geus}, \& {de Zeeuw}}]{brown94}
{Brown}, A.~G.~A., {de Geus}, E.~J., \& {de Zeeuw}, P.~T. 1994, A\&A, 289, 101

\bibitem[{{Burningham} {et~al.}(2005){Burningham}, {Naylor}, {Littlefair}, \&
  {Jeffries}}]{burningham05a}
{Burningham}, B., {Naylor}, T., {Littlefair}, S.~P., \& {Jeffries}, R.~D. 2005,
  MNRAS, 356, 1583

\bibitem[{{Caballero}(2007{\natexlab{a}})}]{caballero07c}
{Caballero}, J.~A. 2007{\natexlab{a}}, Astronomische Nachrichten, 328, 917

\bibitem[{{Caballero}(2007{\natexlab{b}})}]{caballero07b}
{Caballero}, J.~A. 2007{\natexlab{b}}, A\&A, 466, 917

\bibitem[{{Caballero}(2008{\natexlab{a}})}]{caballero08b}
{Caballero}, J.~A. 2008{\natexlab{a}}, MNRAS, 383, 750

\bibitem[{{Caballero}(2008{\natexlab{b}})}]{caballero08a}
{Caballero}, J.~A. 2008{\natexlab{b}}, MNRAS, 383, 375

\bibitem[{{Caballero}(2008{\natexlab{c}})}]{caballero08c}
{Caballero}, J.~A. 2008{\natexlab{c}}, A\&A, 478, 667

\bibitem[{{Caballero} {et~al.}(2007){Caballero}, {B{\'e}jar}, {Rebolo},
  {Eisl{\"o}ffel}, {Zapatero Osorio}, {Mundt}, {Barrado Y Navascu{\'e}s},
  {Bihain}, {Bailer-Jones}, {Forveille}, \& {Mart{\'{\i}}n}}]{caballero07d}
{Caballero}, J.~A., {B{\'e}jar}, V.~J.~S., {Rebolo}, R., {et~al.} 2007, A\&A,
  470, 903

\bibitem[{{Caballero} {et~al.}(2004){Caballero}, {B{\'e}jar}, {Rebolo}, \&
  {Zapatero Osorio}}]{caballero04}
{Caballero}, J.~A., {B{\'e}jar}, V.~J.~S., {Rebolo}, R., \& {Zapatero Osorio},
  M.~R. 2004, A\&A, 424, 857

\bibitem[{{Caballero} {et~al.}(2006{\natexlab{a}}){Caballero}, {Mart{\'{\i}}n},
  {Dobbie}, \& {Barrado Y Navascu{\'e}s}}]{caballero06b}
{Caballero}, J.~A., {Mart{\'{\i}}n}, E.~L., {Dobbie}, P.~D., \& {Barrado Y
  Navascu{\'e}s}, D. 2006{\natexlab{a}}, A\&A, 460, 635

\bibitem[{{Caballero} {et~al.}(2006{\natexlab{b}}){Caballero}, {Mart{\'{\i}}n},
  {Zapatero Osorio}, {B{\'e}jar}, {Rebolo}, {Pavlenko}, \&
  {Wainscoat}}]{caballero06a}
{Caballero}, J.~A., {Mart{\'{\i}}n}, E.~L., {Zapatero Osorio}, M.~R., {et~al.}
  2006{\natexlab{b}}, A\&A, 445, 143

\bibitem[{{Caballero} {et~al.}(2008){Caballero}, {Valdivielso},
  {Mart{\'{\i}}n}, {Montes}, {Pascual}, \&
  {P{\'e}rez-Gonz{\'a}lez}}]{caballero08f}
{Caballero}, J.~A., {Valdivielso}, L., {Mart{\'{\i}}n}, E.~L., {et~al.} 2008,
  A\&A, 491, 515

\bibitem[{{Casali} {et~al.}(2007){Casali}, {Adamson}, {Alves de Oliveira},
  {Almaini}, {Burch}, {Chuter}, {Elliot}, \& {23 co-authors}}]{casali07}
{Casali}, M., {Adamson}, A., {Alves de Oliveira}, C., {et~al.} 2007, A\&A, 467,
  777

\bibitem[{{Chabrier}(2003)}]{chabrier03}
{Chabrier}, G. 2003, PASP, 115, 763

\bibitem[{{Chabrier} {et~al.}(2000){Chabrier}, {Baraffe}, {Allard}, \&
  {Hauschildt}}]{chabrier00c}
{Chabrier}, G., {Baraffe}, I., {Allard}, F., \& {Hauschildt}, P. 2000, ApJ,
  542, 464

\bibitem[{{Cutri} {et~al.}(2003){Cutri}, {Skrutskie}, {van Dyk}, {Beichman},
  {Carpenter}, {Chester}, {Cambresy}, {Evans}, {Fowler}, {Gizis}, \& {15
  coauthors}}]{cutri03}
{Cutri}, R.~M., {Skrutskie}, M.~F., {van Dyk}, S., {et~al.} 2003, 2MASS All Sky
  Catalog of point sources, 2246

\bibitem[{{Dahn} {et~al.}(2002){Dahn}, {Harris}, {Vrba}, {Guetter}, {Canzian},
  {Henden}, {Levine}, {Luginbuhl}, {Monet}, {Monet}, {Pier}, {Stone}, {Walker},
  {Burgasser}, {Gizis}, {Kirkpatrick}, {Liebert}, \& {Reid}}]{dahn02}
{Dahn}, C.~C., {Harris}, H.~C., {Vrba}, F.~J., {et~al.} 2002, AJ, 124, 1170

\bibitem[{{Dobbie} {et~al.}(2002){Dobbie}, {Pinfield}, {Jameson}, \&
  {Hodgkin}}]{dobbie02b}
{Dobbie}, P.~D., {Pinfield}, D.~J., {Jameson}, R.~F., \& {Hodgkin}, S.~T. 2002,
  MNRAS, 335, L79

\bibitem[{{Franciosini} {et~al.}(2006){Franciosini}, {Pallavicini}, \&
  {Sanz-Forcada}}]{franciosini06}
{Franciosini}, E., {Pallavicini}, R., \& {Sanz-Forcada}, J. 2006, A\&A, 446,
  501

\bibitem[{{Garrison}(1967)}]{garrison67}
{Garrison}, R.~F. 1967, \pasp, 79, 433

\bibitem[{{Gatti} {et~al.}(2008){Gatti}, {Natta}, {Randich}, {Testi}, \&
  {Sacco}}]{gatti08}
{Gatti}, T., {Natta}, A., {Randich}, S., {Testi}, L., \& {Sacco}, G. 2008,
  A\&A, 481, 423

\bibitem[{{Gehrels}(1986)}]{gehrels86}
{Gehrels}, N. 1986, ApJ, 303, 336

\bibitem[{{Golimowski} {et~al.}(2004){Golimowski}, {Leggett}, {Marley}, {Fan},
  {Geballe}, {Knapp}, {Vrba}, {Henden}, \& {11 authors}}]{golimowski04a}
{Golimowski}, D.~A., {Leggett}, S.~K., {Marley}, M.~S., {et~al.} 2004, AJ, 127,
  3516

\bibitem[{{Gonz{\'a}lez-Garc{\'{\i}}a}
  {et~al.}(2006){Gonz{\'a}lez-Garc{\'{\i}}a}, {Zapatero Osorio}, {B{\'e}jar},
  {Bihain}, {Barrado Y Navascu{\'e}s}, {Caballero}, \&
  {Morales-Calder{\'o}n}}]{gonzales_garcia06}
{Gonz{\'a}lez-Garc{\'{\i}}a}, B.~M., {Zapatero Osorio}, M.~R., {B{\'e}jar},
  V.~J.~S., {et~al.} 2006, A\&A, 460, 799

\bibitem[{{Hambly} {et~al.}(2008){Hambly}, {Collins}, {Cross}, {Mann}, {Read},
  {Sutorius}, {Bond}, {Bryant}, {Emerson}, {Lawrence}, {Rimoldini}, {Stewart},
  {Williams}, {Adamson}, {Hirst}, {Dye}, \& {Warren}}]{hambly08}
{Hambly}, N.~C., {Collins}, R.~S., {Cross}, N.~J.~G., {et~al.} 2008, MNRAS,
  384, 637

\bibitem[{{Hern{\'a}ndez} {et~al.}(2007){Hern{\'a}ndez}, {Hartmann}, {Megeath},
  {Gutermuth}, {Muzerolle}, {Calvet}, {Vivas}, {Brice{\~n}o}, {Allen},
  {Stauffer}, {Young}, \& {Fazio}}]{hernandez07}
{Hern{\'a}ndez}, J., {Hartmann}, L., {Megeath}, T., {et~al.} 2007, ApJ, 662,
  1067

\bibitem[{{Hewett} {et~al.}(2006){Hewett}, {Warren}, {Leggett}, \&
  {Hodgkin}}]{hewett06}
{Hewett}, P.~C., {Warren}, S.~J., {Leggett}, S.~K., \& {Hodgkin}, S.~T. 2006,
  MNRAS, 367, 454

\bibitem[{{Jameson} {et~al.}(2008){Jameson}, {Lodieu}, {Casewell}, {Bannister},
  \& {Dobbie}}]{jameson08b}
{Jameson}, R.~F., {Lodieu}, N., {Casewell}, S.~L., {Bannister}, N.~P., \&
  {Dobbie}, P.~D. 2008, MNRAS, 385, 1771

\bibitem[{{Jeffries} {et~al.}(2006){Jeffries}, {Maxted}, {Oliveira}, \&
  {Naylor}}]{jeffries06}
{Jeffries}, R.~D., {Maxted}, P.~F.~L., {Oliveira}, J.~M., \& {Naylor}, T. 2006,
  MNRAS, 371, L6

\bibitem[{{Kenyon} {et~al.}(2005){Kenyon}, {Jeffries}, {Naylor}, {Oliveira}, \&
  {Maxted}}]{kenyon05}
{Kenyon}, M.~J., {Jeffries}, R.~D., {Naylor}, T., {Oliveira}, J.~M., \&
  {Maxted}, P.~F.~L. 2005, MNRAS, 356, 89

\bibitem[{{Kharchenko} {et~al.}(2005){Kharchenko}, {Piskunov}, {R{\"o}ser},
  {Schilbach}, \& {Scholz}}]{kharchenko05a}
{Kharchenko}, N.~V., {Piskunov}, A.~E., {R{\"o}ser}, S., {Schilbach}, E., \&
  {Scholz}, R.-D. 2005, A\&A, 438, 1163

\bibitem[{{Kroupa}(2002)}]{kroupa02}
{Kroupa}, P. 2002, Science, 295, 82

\bibitem[{{Lawrence} {et~al.}(2007){Lawrence}, {Warren}, {Almaini}, {Edge},
  {Hambly}, \& {17 co-authors}}]{lawrence07}
{Lawrence}, A., {Warren}, S.~J., {Almaini}, O., {et~al.} 2007, MNRAS, 379, 1599

\bibitem[{{Lee}(1968)}]{lee68}
{Lee}, T.~A. 1968, ApJ, 152, 913

\bibitem[{{Lodieu} {et~al.}(2007{\natexlab{a}}){Lodieu}, {Dobbie}, {Deacon},
  {Hodgkin}, {Hambly}, \& {Jameson}}]{lodieu07c}
{Lodieu}, N., {Dobbie}, P.~D., {Deacon}, N.~R., {et~al.} 2007{\natexlab{a}},
  MNRAS, 380, 712

\bibitem[{{Lodieu} {et~al.}(2006){Lodieu}, {Hambly}, \& {Jameson}}]{lodieu06}
{Lodieu}, N., {Hambly}, N.~C., \& {Jameson}, R.~F. 2006, MNRAS, 373, 95

\bibitem[{{Lodieu} {et~al.}(2008){Lodieu}, {Hambly}, {Jameson}, \&
  {Hodgkin}}]{lodieu08a}
{Lodieu}, N., {Hambly}, N.~C., {Jameson}, R.~F., \& {Hodgkin}, S.~T. 2008,
  MNRAS, 383, 1385

\bibitem[{{Lodieu} {et~al.}(2007{\natexlab{b}}){Lodieu}, {Hambly}, {Jameson},
  {Hodgkin}, {Carraro}, \& {Kendall}}]{lodieu07a}
{Lodieu}, N., {Hambly}, N.~C., {Jameson}, R.~F., {et~al.} 2007{\natexlab{b}},
  MNRAS, 374, 372

\bibitem[{{Low} \& {Lynden-Bell}(1976)}]{low76}
{Low}, C. \& {Lynden-Bell}, D. 1976, MNRAS, 176, 367

\bibitem[{{Lucas} {et~al.}(2008){Lucas}, {Hoare}, {Longmore}, {Schr{\"o}der},
  \& {27 co-authors}}]{lucas08}
{Lucas}, P.~W., {Hoare}, M.~G., {Longmore}, A., {Schr{\"o}der}, A.~C., \& {27
  co-authors}. 2008, MNRAS, 391, 136

\bibitem[{{Luhman}(2004)}]{luhman04a}
{Luhman}, K.~L. 2004, ApJ, 617, 1216

\bibitem[{{Luhman}(2007)}]{luhman07d}
{Luhman}, K.~L. 2007, ApJS, 173, 104

\bibitem[{{Luhman} {et~al.}(2008){Luhman}, {Hern{\'a}ndez}, {Downes},
  {Hartmann}, \& {Brice{\~n}o}}]{luhman08c}
{Luhman}, K.~L., {Hern{\'a}ndez}, J., {Downes}, J.~J., {Hartmann}, L., \&
  {Brice{\~n}o}, C. 2008, ApJ, 688, 362

\bibitem[{{Luhman} {et~al.}(2000){Luhman}, {Rieke}, {Young}, {Cotera}, {Chen},
  {Rieke}, {Schneider}, \& {Thompson}}]{luhman00b}
{Luhman}, K.~L., {Rieke}, G.~H., {Young}, E.~T., {et~al.} 2000, ApJ, 540, 1016

\bibitem[{{Luhman} {et~al.}(2003){Luhman}, {Stauffer}, {Muench}, {Rieke},
  {Lada}, {Bouvier}, \& {Lada}}]{luhman03b}
{Luhman}, K.~L., {Stauffer}, J.~R., {Muench}, A.~A., {et~al.} 2003, ApJ, 593,
  1093

\bibitem[{{Lynga}(1981)}]{lynga81}
{Lynga}, G. 1981, Astronomical Data Center Bulletin, 1, 90

\bibitem[{{Mart{\'{\i}}n} {et~al.}(2001){Mart{\'{\i}}n}, {Zapatero Osorio},
  {Barrado y Navascu{\' e}s}, {B{\' e}jar}, \& {Rebolo}}]{martin01a}
{Mart{\'{\i}}n}, E.~L., {Zapatero Osorio}, M.~R., {Barrado y Navascu{\' e}s},
  D., {B{\' e}jar}, V.~J.~S., \& {Rebolo}, R. 2001, ApJL, 558, L117

\bibitem[{{Maxted} {et~al.}(2008){Maxted}, {Jeffries}, {Oliveira}, {Naylor}, \&
  {Jackson}}]{maxted08}
{Maxted}, P.~F.~L., {Jeffries}, R.~D., {Oliveira}, J.~M., {Naylor}, T., \&
  {Jackson}, R.~J. 2008, MNRAS, 385, 2210

\bibitem[{{Miller} \& {Scalo}(1979)}]{miller79}
{Miller}, G.~E. \& {Scalo}, J.~M. 1979, ApJS, 41, 513

\bibitem[{{Oliveira} {et~al.}(2002){Oliveira}, {Jeffries}, {Kenyon},
  {Thompson}, \& {Naylor}}]{oliveira02}
{Oliveira}, J.~M., {Jeffries}, R.~D., {Kenyon}, M.~J., {Thompson}, S.~A., \&
  {Naylor}, T. 2002, A\&A, 382, L22

\bibitem[{{Perryman} {et~al.}(1997){Perryman}, {Lindegren}, {Kovalevsky},
  {Hoeg}, \& {15 co-authors}}]{perryman97}
{Perryman}, M.~A.~C., {Lindegren}, L., {Kovalevsky}, J., {Hoeg}, E., \& {15
  co-authors}. 1997, A\&A, 323, L49

\bibitem[{{Sacco} {et~al.}(2008){Sacco}, {Franciosini}, {Randich}, \&
  {Pallavicini}}]{sacco08}
{Sacco}, G.~G., {Franciosini}, E., {Randich}, S., \& {Pallavicini}, R. 2008,
  A\&A, 488, 167

\bibitem[{{Salpeter}(1955)}]{salpeter55}
{Salpeter}, E.~E. 1955, ApJ, 121, 161

\bibitem[{{Scalo}(1986)}]{scalo86}
{Scalo}, J.~M. 1986, Fundamentals of Cosmic Physics, 11, 1

\bibitem[{{Scholz} \& {Jayawardhana}(2008)}]{scholz08a}
{Scholz}, A. \& {Jayawardhana}, R. 2008, ApJL, 672, L49

\bibitem[{{Sherry} {et~al.}(2004){Sherry}, {Walter}, \& {Wolk}}]{sherry04}
{Sherry}, W.~H., {Walter}, F.~M., \& {Wolk}, S.~J. 2004, AJ, 128, 2316

\bibitem[{{Tokunaga} {et~al.}(2002){Tokunaga}, {Simons}, \&
  {Vacca}}]{tokunaga02}
{Tokunaga}, A.~T., {Simons}, D.~A., \& {Vacca}, W.~D. 2002, PASP, 114, 180

\bibitem[{{Walter} {et~al.}(1994){Walter}, {Vrba}, {Mathieu}, {Brown}, \&
  {Myers}}]{walter94}
{Walter}, F.~M., {Vrba}, F.~J., {Mathieu}, R.~D., {Brown}, A., \& {Myers},
  P.~C. 1994, AJ, 107, 692

\bibitem[{{Weaver} \& {Babcock}(2004)}]{weaver04}
{Weaver}, W.~B. \& {Babcock}, A. 2004, PASP, 116, 1035

\bibitem[{{West} {et~al.}(2008){West}, {Hawley}, {Bochanski}, {Covey}, {Reid},
  {Dhital}, {Hilton}, \& {Masuda}}]{west08}
{West}, A.~A., {Hawley}, S.~L., {Bochanski}, J.~J., {et~al.} 2008, AJ, 135, 785

\bibitem[{{Whitworth} \& {Zinnecker}(2004)}]{whitworth04}
{Whitworth}, A.~P. \& {Zinnecker}, H. 2004, A\&A, 427, 299

\bibitem[{{Zapatero Osorio} {et~al.}(2002{\natexlab{a}}){Zapatero Osorio},
  {B{\' e}jar}, {Mart{\'{\i}}n}, {Barrado y Navascu{\' e}s}, \&
  {Rebolo}}]{zapatero02a}
{Zapatero Osorio}, M.~R., {B{\' e}jar}, V.~J.~S., {Mart{\'{\i}}n}, E.~L.,
  {Barrado y Navascu{\' e}s}, D., \& {Rebolo}, R. 2002{\natexlab{a}}, ApJL,
  569, L99

\bibitem[{{Zapatero Osorio} {et~al.}(2000){Zapatero Osorio}, {B{\' e}jar},
  {Mart{\'{\i}}n}, {Rebolo}, {Barrado y Navascu{\' e}s}, {Bailer-Jones}, \&
  {Mundt}}]{zapatero00}
{Zapatero Osorio}, M.~R., {B{\' e}jar}, V.~J.~S., {Mart{\'{\i}}n}, E.~L.,
  {et~al.} 2000, Science, 290, 103

\bibitem[{{Zapatero Osorio} {et~al.}(2002{\natexlab{b}}){Zapatero Osorio},
  {B{\' e}jar}, {Pavlenko}, {Rebolo}, {Allende Prieto}, {Mart{\'{\i}}n}, \&
  {Garc{\'{\i}}a L{\' o}pez}}]{zapatero02c}
{Zapatero Osorio}, M.~R., {B{\' e}jar}, V.~J.~S., {Pavlenko}, Y., {et~al.}
  2002{\natexlab{b}}, A\&A, 384, 937

\bibitem[{{Zapatero Osorio} {et~al.}(2007){Zapatero Osorio}, {Caballero},
  {B{\'e}jar}, {Rebolo}, {Barrado Y Navascu{\'e}s}, {Bihain}, {Eisl{\"o}ffel},
  {Mart{\'{\i}}n}, {Bailer-Jones}, {Mundt}, {Forveille}, \&
  {Bouy}}]{zapatero07b}
{Zapatero Osorio}, M.~R., {Caballero}, J.~A., {B{\'e}jar}, V.~J.~S., {et~al.}
  2007, A\&A, 472, L9

\end{thebibliography}

%
%
\appendix

%
%
%
\section{Proper motion non members and possible binaries}
\begin{table*}
 \centering
  \caption{Near-infrared photometry and proper motions
for five sources rejected as proper motion members and three potential
binary previously unresolved on the 2MASS images.
}
 \label{tab_sOri_GCS:GCScand_PM_NM}
 \begin{tabular}{l c c c c c c c l}
 \hline
IAU Names                     &  $Z$   &   $Y$  &   $J$  &   $H$  &  $K$   &  $\mu_{\alpha}\cos{\delta}$ & $\mu_{\delta}$ & Comments \\
                              &  mag   &   mag  &   mag  &   mag  &  mag   &    mas/yr                  &  mas/yr                    \\
\hline
UGCS J053703.74$-$024150.0 & 14.943 & 14.453 & 13.894 & 13.329 & 13.038 & $-$3.7  & $-$36.5 & PM NM \\
UGCS J053855.48$-$024129.6 & 12.922 & 12.606 & 12.139 & 11.705 & 11.342 &    91.1 &    13.1 & PM NM \\
UGCS J053856.65$-$025702.3 & 12.801 & 12.531 & 12.062 & 11.723 & 11.309 &    33.8 & $-$14.9 & PM NM \\
UGCS J053907.60$-$022905.6 & 16.201 & 15.553 & 14.874 & 14.299 & 13.928 &    43.3 &  $-$7.0 & PM NM \\
UGCS J054019.78$-$022956.0 & 13.285 & 12.947 & 12.453 & 11.945 & 11.599 &    39.5 & $-$23.4 & PM NM \\
 \hline
UGCS J053847.90$-$023719.4 & 13.148 & 12.771 & 12.205 & 11.449 & 10.908 &  $-$28.8 &  $-$26.3 & Binary? \\
UGCS J053848.03$-$023718.5 & 13.485 & 13.152 & 12.673 & 11.997 & 11.736 &    242.7 &    107.6 & Binary? \\
UGCS J053947.30$-$022618.3 & 13.493 & 13.113 & 12.605 & 11.969 & 11.718 & $-$249.6 & $-$291.6 & Binary? \\
 \hline
 \end{tabular}
\begin{list}{}{}
\item[$^{a}$] IAU (International Astronomical Union) designations include J2000 (RA,dec) coordinates from UKIDSS GCS DR4 
(see http://www.ukidss.org/archive/archive.html for the UKIDSS nomenclature).
\end{list}
\end{table*}

%
%
%
%
%
\section{Faint members identified in GCS DR4}
\begin{table*}
 \centering
  \caption{Near-infrared photometry and its associated errors for 10 faint
candidate members identified in $YJHK$ (top) and 5 in $JHK$ (bottom).}
 \label{tab_sOri_GCS:new_FAINT}
 \begin{tabular}{c c c c c c l}
 \hline
IAU Name$^{a}$                 &  $Y$  &   $J$  &   $H$  &  $K$   & Comments \\
                               &  mag  &   mag  &   mag  &   mag  &          \\
 \hline
UGCS J053957.39$-$025006.1  & 20.339$\pm$0.276  & 18.797$\pm$0.104  & 17.950$\pm$0.116  & 17.349$\pm$0.095  &            \\
UGCS J053858.55$-$025226.6  & 20.137$\pm$0.219  & 18.677$\pm$0.089  & 17.990$\pm$0.113  & 17.196$\pm$0.079  &            \\
UGCS J053900.79$-$022141.8  & 19.967$\pm$0.201  & 18.360$\pm$0.073  & 17.783$\pm$0.107  & 16.867$\pm$0.065  & SOri56     \\
UGCS J053824.71$-$030028.3  & 20.099$\pm$0.193  & 18.825$\pm$0.098  & 17.805$\pm$0.107  & 17.357$\pm$0.109  &            \\
UGCS J053829.50$-$025959.1  & 19.498$\pm$0.113  & 18.425$\pm$0.069  & 17.844$\pm$0.111  & 16.935$\pm$0.074  &            \\
UGCS J053713.22$-$022449.1  & 19.904$\pm$0.148  & 18.783$\pm$0.090  & 17.417$\pm$0.048  & 17.357$\pm$0.093  & edge       \\
UGCS J053916.72$-$023348.4  & 20.028$\pm$0.192  & 18.950$\pm$0.111  & 18.602$\pm$0.195  & 17.405$\pm$0.095  &            \\
UGCS J053937.52$-$023041.9  & 20.147$\pm$0.208  & 18.961$\pm$0.110  & 18.199$\pm$0.130  & 17.622$\pm$0.114  & SOri60     \\
UGCS J053944.98$-$023818.7  & 20.174$\pm$0.220  & 18.950$\pm$0.111  & 17.734$\pm$0.087  & 16.958$\pm$0.063  & dubious    \\
UGCS J053949.52$-$023129.7  & 20.087$\pm$0.197  & 18.922$\pm$0.106  & 18.291$\pm$0.140  & 17.502$\pm$0.102  & red        \\
 \hline
UGCS J053857.52$-$022905.5  &  ---              & 18.963$\pm$0.181  & 18.472$\pm$0.101  & 17.483$\pm$0.000  &            \\
UGCS J053951.31$-$024942.1  &  ---              & 17.895$\pm$0.054  & 17.154$\pm$0.054  & 16.787$\pm$0.000  & close to bright star    \\
UGCS J053911.32$-$022431.2  &  ---              & 18.968$\pm$0.142  & 18.108$\pm$0.099  & 17.337$\pm$0.000  & in Subaru survey    \\
UGCS J053834.00$-$025753.3  &  ---              & 17.599$\pm$0.054  & 17.044$\pm$0.051  & 16.511$\pm$0.000  & close to bright star   \\
UGCS J053904.73$-$020831.5  &  ---              & 18.979$\pm$0.121  & 17.971$\pm$0.113  & 17.451$\pm$0.000  & in Subaru survey     \\
 \hline
 \end{tabular}
\begin{list}{}{}
\item[$^{a}$] IAU designations include J2000 (RA,dec) coordinates from UKIDSS GCS DR4\@.
\end{list}
\end{table*}

\section{Members recovered by the GCS}
\begin{table*}
 \centering
  \caption{Near-infrared photometry and its associated
errors from the fourth data release of the GCS for previously known 
members of \sOri{}.}
 \label{tab_sOri_GCS:Known_Memb}
 \begin{tabular}{c c c c c c c c c c}
 \hline
R.A.$^{a}$  &  Dec.$^{a}$  & $Z$  &  $Y$  &  $J$  &  $H$  & $K$ &  $\mu_{\alpha}\cos{\delta}$ & $\mu_{\delta}$ & Comments$^{b}$ \\
 h:m:s      &   d:':''            &  mag  &  mag  &  mag  & mag &     mas/yr                  &  mas/yr        &                \\
 \hline
05:36:46.91 & $-$02:33:28.3 & 14.636$\pm$0.003  & 14.086$\pm$0.002  & 13.474$\pm$0.002  & 12.927$\pm$0.002  & 12.644$\pm$0.002  &     8.1 &   $-$5.5 & M1773275 \\
05:36:58.07 & $-$02:35:19.4 & 16.457$\pm$0.009  & 15.691$\pm$0.006  & 15.043$\pm$0.006  & 14.540$\pm$0.005  & 14.174$\pm$0.006  &  $-$7.3 &   $-$3.0 & M1599271 \\
\ldots{}    & \ldots{}     & \ldots{}          & \ldots{}          & \ldots{}          & \ldots{}          & \ldots{}          & \ldots{}  & \ldots{} & \ldots{} \\
05:40:09.32 & $-$02:26:32.6 & 19.484$\pm$0.085  & 18.304$\pm$0.039  & 17.392$\pm$0.027  & 16.878$\pm$0.039  & 16.209$\pm$0.032  & --- & --- & SOri\,52  \\
05:40:13.96 & $-$02:31:27.4 & 16.295$\pm$0.008  & 15.272$\pm$0.005  & 14.492$\pm$0.004  & 14.034$\pm$0.004  & 13.560$\pm$0.003  &    5.9 &   $-$4.1 & KJN63      \\
05:40:34.40 & $-$02:44:09.5 & 14.951$\pm$0.004  & 14.291$\pm$0.003  & 13.681$\pm$0.003  & 13.155$\pm$0.002  & 12.777$\pm$0.002  & $-$5.9 &  8.3 & BNLJ63  \\
 \hline
 \end{tabular}
\begin{list}{}{}
\item[$^{a}$] Coordinates in J2000 from the UKIDSS GCS DR4\@.
\item[$^{b}$] Names in the last column are as follows: M stands for the Mayrit catalogue \citep{caballero08c}, KJN for members in \citet{kenyon05}, BNLJ for objects in Table 3 of \citet{burningham05a}, and SOri nomenclature comes from the work by \citet{bejar99}, \citet{zapatero00}, and \citet{bejar01}.
Four sources are spectroscopic binaries \citep[M258337, M459224, M873229, M1493050;][]{maxted08} and three (S36, S42, S97) are classified as non-members by \citet{sacco08}.
\end{list}
\end{table*}
\section{Published sources classified as non-members by the GCS}
\begin{table*}
 \centering
  \caption{Photometry and proper motions of known 
sources non selected as potential member candidates by the GCS\@.}
 \label{tab_sOri_GCS:NM}
 \begin{tabular}{c c c c c c c c c l}
 \hline
R.A.$^{a}$       & Dec.$^{a}$        &  $Z$   &   $Y$  &   $J$  &   $H$  &  $K$   & $\mu_{\alpha}\cos{\delta}$ & $\mu_{\delta}$ & Comments$^{b}$ \\
 h:m:s           &   d:':''          &  mag   &   mag  &   mag  &   mag  &  mag   &    mas/yr                  &  mas/yr        &                \\
\hline
05:38:38.89 & $-$02:28:01.7 & 16.332 & 15.844 & 15.276 & 14.735 & 14.405 &  $-$0.4 & $-$10.3 & M487350 ([SE 2004]70; blue $Z-J$) ) \\
05:38:41.46 & $-$02:35:52.3 & 15.478 & 99.999 & 13.963 & 13.399 & 12.958 &  $-$0.5 &  $-$0.8 & M500279; No $Y$ photometry \\
05:38:47.90 & $-$02:37:19.4 & 13.148 & 12.771 & 12.205 & 11.449 & 10.908 &  $-$28.8 &  $-$26.3 & M92149 (PM NM) \\
05:38:51.00 & $-$02:49:13.9 & 16.149 & 15.620 & 15.044 & 14.424 & 14.106 &  $-$7.6 &   13.1 & M799173 (blue $Z-J$)  \\
05:39:06.97 & $-$02:12:16.9 & 14.358 & 14.132 & 13.666 & 13.102 & 12.967 &    3.2 &   $-$3.4 & M1462013 (blue $Z-J$)  \\
05:39:07.60 & $-$02:29:05.6 & 16.201 & 15.553 & 14.874 & 14.299 & 13.928 &   43.3 &  $-$7.0 &  M537040 (SOri\,20; PM NM) \\
05:39:26.47 & $-$02:26:15.5 & 14.255 & 13.927 & 13.375 & 12.719 & 12.459 &  $-$0.8 &    0.3 & M856047 (blue $Z-J$) \\
05:40:25.80 & $-$02:48:55.3 & 16.285 & 15.974 & 15.455 & 14.435 & 13.334 &    0.5 &    3.1 & M1701117 (blue $Z-J$) \\
\hline
05:39:16.58 & $-$02:38:25.4 & 20.270 & 19.837 & 19.113 & 18.204 & 18.151 &  ---    &  ---    &  SOri\,26 (M4.5) \\
05:37:07.21 & $-$02:32:44.2 & 16.649 & 16.098 & 15.498 & 14.937 & 14.614 &  7.5    & 12.8    &  SOri\,34     \\
05:39:38.50 & $-$02:31:13.3 & 17.870 & 17.285 & 16.590 & 16.061 & 15.719 &  ---    &  ---    &  SOri\,41     \\
05:38:13.96 & $-$02:35:01.3 & 18.335 & 17.591 & 16.926 & 16.376 & 15.966 &  ---    &  ---    &  SOri\,43     \\
05:38:07.11 & $-$02:43:20.9 & 18.602 & 17.882 & 17.172 & 16.667 & 16.244 &  ---    &  ---    &  SOri\,44 (M7)  \\
05:38:23.02 & $-$02:37:55.1 & 99.999 & 19.670 & 18.849 & 18.031 & 17.862 &  ---    &  ---    &  SOri\,49 (M7.5) \\
05:39:47.05 & $-$02:25:24.5 & 99.999 & 19.988 & 18.934 & 99.999 & 17.932 &  ---    &  ---    &  SOri\,57      \\
05:39:10.02 & $-$02:28:11.5 & 15.504 & 15.044 & 14.533 & 13.990 & 13.706 &  20.6   & 23.6    &  SOri J053909.9-022814 (M5) \\
 \hline
05:40:34.61 & $-$02:33:13.8 & 15.103 & 14.675 & 14.115 & 13.489 & 13.204 & $-$6.6   &   12.5  & KJN29, phot NM      \\
05:39:10.02 & $-$02:28:11.5 & 15.504 & 15.044 & 14.533 & 13.990 & 13.706 &   20.6   &   23.6  & KJN33,SOriJ053909.9$-$022814, phot NM \\
05:38:18.17 & $-$02:43:34.9 & 15.473 & 15.137 & 14.636 & 14.036 & 13.775 &    6.6   &    7.6  & KJN34, phot NM      \\
05:40:17.06 & $-$02:26:49.0 & 15.885 & 15.514 & 14.963 & 14.413 & 14.104 &   19.1   & $-$5.7  & KJN45, phot NM      \\
05:38:46.59 & $-$02:19:40.4 & 16.356 & 15.914 & 15.352 & 14.795 & 14.479 &    7.3   &   10.6  & KJN53, phot NM      \\
05:39:43.39 & $-$02:53:23.0 & 16.439 & 15.899 & 15.293 & 14.791 & 14.452 &   12.4   & $-$10.7 & KJN59, phot NM      \\
 \hline
05:38:33.34 & $-$02:20:59.4 & 19.968 & 18.905 & 18.026 & 17.244 & 16.598 &  ---  &  ---  &  SOri\,54 (M9.5) \\
05:39:03.60 & $-$02:25:36.6 & 20.469 & 19.062 & 18.362 & 17.900 & 17.067 &  ---  &  ---  &  SOri\,58 (L0)  \\
05:39:26.77 & $-$02:26:14.3 & 17.633 & 16.888 & 16.224 & 15.688 & 15.261 &  ---  &  ---  &  SOri J053926.8-022614   \\
05:39:48.26 & $-$02:29:14.3 & 17.881 & 17.079 & 16.382 & 15.834 & 15.366 &  ---  &  ---  &  SOri J053948.1-022914 (M7) \\
 \hline
05:37:18.69 & $-$02:40:22.0 & 13.103 & 12.807 & 12.314 & 12.055 & 11.800 & $-$17.4 &  $-$25.7 & SO\,9    \\
05:37:23.05 & $-$02:32:46.7 & 15.126 & 99.999 & 14.159 & 13.592 & 13.319 &  $-$22.3 & $-$14.2 & SO\,27   \\
\ldots{}    & \ldots{}     & \ldots{} & \ldots{} & \ldots{} & \ldots{} & \ldots{}  & \ldots{}  & \ldots{} & \ldots{} \\
05:39:56.83 & $-$02:53:14.6 & 20.433 & 19.160 & 18.202 & 17.746 & 17.183 &     ---  &     ---  &  \\
05:39:57.62 & $-$02:47:36.0 & 13.995 & 13.668 & 13.165 & 12.537 & 12.326 &  $-$7.3 &  $-$4.7 & SO\,1296 \\
 \hline
 \end{tabular}
\begin{list}{}{}
\item[$^{a}$] Coordinates in J2000 from UKIDSS GCS DR4
\item[$^{b}$] The last column provides the original names found in the literature.
\item[Note:] The first part of the table lists objects from the Mayrit catalogue; the second part lists objects published by \citet{bejar99}, \citet{zapatero00}, and \citet{bejar01}, the third part lists candidates classified as ``Maybe'' by \citet{kenyon05}, the fourth part lists objects lying at the borderline between members and non-members, the last section gives mid-infrared sources published by \citet{hernandez07} and \citet{luhman08c}.
\end{list}
\end{table*}

%
%

%
%
%
\section{New members identified in the GCS}
\begin{table*}
 \centering
  \caption{Near-infrared and proper motions 
for the new candidate members$^{a}$ of \sOri{} identified in the GCS $^{b}$.}
 \label{tab_sOri_GCS:new_MEMB}
 \begin{tabular}{l c c c c c c c}
 \hline
IAU Names$^{c}$               &  $Z$   &   $Y$  &   $J$  &   $H$  &  $K$   &  $\mu_{\alpha}\cos{\delta}$ & $\mu_{\delta}$ \\
                              &  mag   &   mag  &   mag  &   mag  &  mag   &    mas/yr                  &  mas/yr          \\
\hline
UGCS J053649.28$-$024354.4 & 14.363$\pm$0.003  & 13.876$\pm$0.002  & 13.315$\pm$0.002  & 12.666$\pm$0.002  & 12.352$\pm$0.001  &  2.1 & -1.3 \\
UGCS J053649.97$-$023522.6 & 14.931$\pm$0.004  & 14.424$\pm$0.003  & 13.849$\pm$0.003  & 13.259$\pm$0.002  & 13.003$\pm$0.002  & $-$3.6 & -1.2 \\
\ldots{}                      & \ldots{}          & \ldots{}          & \ldots{}          & \ldots{}          & \ldots{}          & \ldots{}  & \ldots{} \\
UGCS J054020.79$-$022400.1 & 18.688$\pm$0.044  & 17.666$\pm$0.023  & 16.886$\pm$0.018  & 16.377$\pm$0.025  & 15.789$\pm$0.022  & --- & --- \\
UGCS J054034.08$-$022602.5 & 18.042$\pm$0.026  & 17.331$\pm$0.018  & 16.515$\pm$0.014  & 15.756$\pm$0.015  & 15.230$\pm$0.014  & --- & --- \\
 \hline
 \end{tabular}
\begin{list}{}{}
\item[$^{a}$] Additional $IzJ$ ($HK$) photometry is available in \citet{bejar01_PhD}.
\item[$^{b}$] The full list is ordered by right ascension and is available in the electronic version of the journal 
\item[$^{c}$] IAU designations include J2000 (RA,dec) coordinates from UKIDSS GCS DR4
\end{list}
\end{table*}

%
%
%
%
\section{Variable sources}
\begin{table*}
 \centering
  \caption{Near-infrared photometry from the UKIDSS GCS DR4
compared to 2MASS for variable sources (with 99.5\%
confidence level).
}
 \label{tab_sOri_GCS:GCScand_variable}
 \begin{tabular}{l c c c c c c c c c l}
 \hline
IAU Name$^{a}$                &  $J$  &   $H$  &  $K$   &  $J_{\rm 2MASS}$ & $H_{\rm 2MASS}$ & $K_{\rm 2MASS}$ & diff $J$ & diff $H$ & diff $K$ & Comments \cr
                              &  mag  &   mag  &  mag  &      mag          &   mag           &    mag          &   mag    &   mag    &   mag    &       \\
 \hline
UGCS J053753.98$-$024954.5 & 14.076 & 12.786 & 11.960 & 14.520 & 13.250 & 12.455 & $-$0.444 & $-$0.464 & $-$0.495 & M1129222$^{b}$  \\
UGCS J053820.50$-$023408.9 & 13.651 & 12.591 & 11.931 & 12.652 & 11.918 & 11.648 &    0.999 &    0.673 &    0.283 & M380287$^{c}$   \\
UGCS J053834.46$-$025351.5 & 15.292 & 14.262 & 12.973 & 15.535 & 14.043 & 12.651 & $-$0.243 &    0.219 &    0.322 & M1082188$^{d}$  \\
UGCS J053849.29$-$022357.6 & 14.536 & 14.115 & 13.524 & 14.362 & 13.699 & 13.197 &    0.174 &    0.416 &    0.327 & M726005$^{e}$  \\
UGCS J053904.59$-$024149.2 & 13.746 & 13.147 & 12.389 & 13.962 & 12.910 & 12.224 & $-$0.216 &    0.237 &    0.165 & M458140$^{f}$ \\
 \hline
UGCS J053838.59$-$030220.1 & 13.913 & 13.187 & 12.660 & 13.711 & 12.962 & 12.566 &    0.202 &    0.225 &    0.094 & M1583183$^{d}$  \\
UGCS J053855.42$-$024120.9 & 15.336 & 14.615 & 13.979 & 15.622 & 14.842 & 13.968 & $-$0.286 & $-$0.227 &    0.011 & M358154$^{g}$   \\
UGCS J053825.43$-$024241.2 & 14.793 & 14.334 & 13.729 & 14.877 & 14.157 & 13.572 & $-$0.084 &    0.177 &    0.157 & M495216$^{h}$    \\
UGCS J053823.34$-$022534.6 & 13.802 & 13.115 & 12.583 & 13.685 & 12.928 & 12.424 &    0.117 &    0.187 &    0.159 & M703333$^{g}$  \\
UGCS J053838.12$-$023202.6 & 14.808 & 14.270 & 13.449 & 14.800 & 13.771 & 13.197 &    0.008 &    0.499 &    0.252 & M258337$^{g}$   \\
 \hline
 \end{tabular}
\begin{list}{}{}
\item[Note:] The top part of the table lists sources showing variability in $JHK$ whereas the bottom part lists
objects variable in two bands only.
\item[$^{a}$] IAU designations include J2000 (RA,dec) coordinates from UKIDSS GCS DR4\@
\item[$^{b}$] Weak H$\alpha$ detection reported by \citet{weaver04}
\item[$^{c}$] M4 with detection of lithium and H$\alpha$ equivalent width of 28\AA{} \citep{zapatero02c}
\item[$^{d}$] M1082188 and M1583183 were reported for the first time by \citet{caballero08c} as photometric member candidates
\item[$^{e}$] M5 with lithium, H$\alpha$ emission and disk \citep{kenyon05}; no Pa$\beta$/Pa$\gamma$ but HeI detected \citep{gatti08}. Faint companion detected on the GCS images.
\item[$^{f}$] M2 with disk, X-ray emission, HeI, lithium but no Pa$\beta$/Pa$\gamma$ \citep{sacco08,gatti08}. Faint companion detected on the GCS images.
\item[$^{g}$] Presence of disk reported by \citet{hernandez07} and \citet{caballero08c}
\item[$^{h}$] M6, H$\alpha$ emission and presence of disk. Reported to be variable by \citet{caballero04} with detailed discussion in \citet{caballero06a}
\end{list}
\end{table*}

\end{document}